# Nanoparticles of the giant dielectric material , $CaCu_3Ti_4O_{12}$ from a precursor route.

**P.Thomas, [a,b] K.Dwarakanath,[a] K.B.R.Varma[b] , T.R.N.Kutty. [b*]**

[a] Dielectric Materials Division, Central Power Research Institute, Bangalore : 560 080, India.

[b] Materials Research Centre, Indian Institute of Science, Bangalore: 560012,India.

**Abstract:**

A method of preparing the nanoparticles of $CaCu_3Ti_4O_{12}$ (CCTO) with the crystallite size varying from 30-200 nm is optimized at a temperature as low as 680$^o$C from the exothermic thermal decomposition of an oxalate precursor, $CaCu_3(TiO)_4(C_2O_4)_8 \bullet 9H_2O$. The phase singularity of the complex oxalate precursor is confirmed by the wet chemical analyses, X-ray diffraction, FT-IR and TGA/DTA analyses. The UV-VIS reflectance and ESR spectra of CCTO powders indicate that the Cu (II) coordination changes from distorted octahedra to nearly flattened tetrahedra (squashed) to square planar geometry with increasing annealing temperature. The HRTEM images have revealed that the evolution of the microstructure in nanoscale is related to the change in Cu (II) coordination around the surface regions for the chemically prepared powder specimens. The nearly flattened tetrahedral geometry prevails for $CuO_4$ in the near surface regions of the particles whereas square planar $CuO_4$ groups are dominant in the interior regions of the nanoparticles. The powders derived from the oxalate precursor have excellent sinterability resulting in high density ceramics which exhibited giant dielectric constants upto 40,000 (1 kHz) at 25$^o$C, accompanied by low dielectric loss < 0.07.

**Keywords:** A. Oxides; B. Chemical synthesis; C. Electron microscopy ; D. Dielectric properties.

---

* Corresponding author: E-mail: <kutty@mrc.iisc.ernet.in> Tel. +91-80-2293-2784; Fax: +91-80-2360-7316.

## 1. INTRODUCTION

Electroceramics associated with giant dielectric constants have been in increasing demand owing to the miniaturization of electronic devices. Oxides with the perovskite and related structures are well known for their high dielectric constants. The titanate compound $CaCu_3Ti_4O_{12}$ (CCTO) belongs to a family of the type, $ACu_3Ti_4O_{12}$ (where A= Ca or Cd) and has been reported in the year 1967 [1]. This composition has been extended [2] and the family of titanates with the body centered cubic (bcc) structure has been reported having the general formula, $[AC_3]$ $(B_4)$ $O_{12}$, [where A= Ca,Cd,Sr,Na or Th; B = Ti or (Ti + $M^{5+}$), in which M= Ta, Sb or Nb; and C= $Cu^{2+}$or $Mn^{3+}$]. These solids have been synthesized by the ceramic methods and their accurate crystal structures (space group Im3, lattice parameter a ≈ 7.391 Å, and Z=2) determined from the neutron diffraction data. In $CaCu_3Ti_4O_{12}$, $Ca^{2+}$ is dodecahedrally coordinated by oxygen ions, while $Cu^{2+}$ is in square planar coordination (oxygen as the nearest neighbour) and $Ti^{4+}$ coordinates six oxygen ions in a slightly distorted octahedron. The $TiO_6$ groups are tilted by about $20^o$ with respect to the unit cell axis [2]. The crystal structure of $CaCu_3Ti_4O_{12}$ was further refined[2] and found to remain centrosymmetric body centered cube (bcc) over a wide range of temperatures. CCTO has attracted considerable attention recently due to its unusually high dielectric constant ($\varepsilon \sim 10^{4-5}$) which is nearly independent of frequency (upto 10 MHz) and low thermal coefficient of permittivity (TCK) over a wider range of temperature (100-600K) [3, 4]. Several explanations for the origin of high dielectric constant for this solid have been proposed mainly arising from the bulk property contributions as against the microstructural features which in turn are affected by the ceramic processing conditions including sintering temperatures as well as the atmosphere [3,5-7]. Accordingly, the giant dielectric constants have been variously attributed to: (i) the barrier layer capacitance arising at twin boundaries [3]; (ii) disparity in electrical properties between grain interiors and grain boundaries [8-10]; (iii) space charge at the interfaces between the sample and the electrode contacts [11,12]; (iv) polarizability contributions from lattice distortions [13]; (v) differences in electrical properties due to internal domains [14]; (vi) dipolar contributions from oxygen vacancies [15-16]; (vii) the role of Cu off-stoichiometry in modifying the polarization mechanisms [17] ; (viii) cation disorder induced planar defects and associated inhomogeneity [18] or (ix) nanoscale disorder of Ca/Cu substitution giving rise to electronic contribution from the degenerate $e_g$ states of Cu occupying the Ca site contributing to the high dielectric constant [19]. Though several explanations have been put forward, the actual mechanism of the origin of giant dielectric constants in CCTO is still debated as to whether it is intrinsic or extrinsic in nature.

CCTO has been prepared by the solid-solid reactions between the stoichiometric starting materials of $CaCO_3$,$TiO_2$ and CuO. The mixtures are calcined at high temperature for long durations (typically 1000-1050$^o$C for 24-48h) with repeated intermediate grindings [5-8]. This method of preparation is very cumbersome, often requiring temperatures approaching the melting point of CuO. Besides, this procedure suffers from the disadvantages of chemical inhomogeneity leading to coarse particle size for the product. In contrast, the wet chemical syntheses routes offer homogeneous products at lower temperatures in shorter durations. There exist a few research papers indicating that CCTO can be prepared by routes other than the solid state reactions [20-23]. However, these methods [22,23] are complex in nature by way of yielding multiphase products and required longer heat treatments to obtain the phase singularity. The powders prepared by the pyrolysis of the co-precipitated oxalates [24] at 900$^o$C for 10h yielded CCTO with $CaTiO_3$ + CuO as the impurity phases. The phase-pure CCTO was obtained only after sintering the powders at 1050$^o$C. To avoid such difficulties in obtaining phase-pure CCTO powders at relatively lower temperatures, we have presently developed a precursor oxalate route. This is a convenient method for achieving chemical homogeneity where the individual constituents intermix at the ionic level under controlled wet chemical conditions. The phase-pure CCTO nanoparticles have been realized from this precursor route.

## 2. EXPERIMENTAL

### 2.1. Preparation of the oxalate precursor complex

For the precipitation of single-phase calcium copper titanyl oxalate precursor complex (CCT-OX), titania gel was prepared from aqueous $TiOCl_2$ which, in turn, was obtained from the controlled reaction of ice-cold distilled water with $TiCl_4$ (titanium tetrachloride, 99.98%) (Merck, Germany). The other chemicals employed were calcium carbonate/calcium chloride (BDH; A.R.grade), cupric chloride/cupric nitrate (Fluka, proanalyse grade), oxalic acid (S.D. Fine Chemicals, analytical grade), ethanol or acetone (Nice, India; 99.5% pure). As the first step, titania gel [$TiO_2 . xH_2O$ (92<x<118)] (0.4 moles) was prepared from the aqueous $TiOCl_2$(0.05M) by adding $NH_4OH$ (aq) (at 25°C) till the pH reaches ~ 8.0 and washed off $NH_4Cl$ on the filter funnel. This gel was added to 0.4 or 0.8 moles of oxalic acid (2 M solution) (1:1 or 1:2 ratio of Ti: $C_2O_4^{2-}$) which was kept warm at ~40°C. A clear solution was obtained on standing for several hours according to the reactions [Scheme-I]. The different methods attempted for the preparation of complex oxalate precursor using this clear solution are shown in the Scheme – I.

The precipitates obtained from reaction schemes A to C were found to be copper deficient due to the formation of double salts with the neutralizing agents which in turn dissolved partly in water+ethanol. It has been reported in Ref [25] that the aqueous solution containing titanyl oxalic acid + calcium tiatanyl oxalate remained clear without any precipitate formation (reaction Scheme C). This solution was cooled to $10^oC$ to which cupric chloride (or cupric nitrate) dissolved in acetone+water (80/20) was added and stirred continuously. The thick precipitate was separated out by further addition of acetone (reaction scheme D). The precipitate was filtered, washed several times with acetone to make it chloride-free and dried in air. The precipitate obtained by this method of preparation after drying remained finely divided with minimum level of agglomeration. The method was further extended by employing calcium chloride+cupric chloride (or cupric nitrate), wherein the precipitate was more agglomerated, although the products formed by both the schemes were compositionally the same.

The wet chemical analyses of this air-dried precipitate (CCT-OX) gave: Ca: 2.98; Cu: 14.02; TiO: 19.06; $C_2O_4$: 52.08; $H_2O$: 11.87 % (by wt) : Cal. for $CaCu_3(TiO)_4(C_2O_4)_8 \cdot 9H_2O$: Ca: 2.96; Cu: 13.99; TiO: 18.95; $C_2O_4$: 52.10; $H_2O$: 11.99 %:

**2.2. Characterization Techniques**

X-ray powder diffraction studies were carried out with an X'PERT-PRO Diffractometer (Philips, Netherlands) using Cu $K\alpha_1$ radiation ($\lambda$ = 0.154056 nm) in a wide range of $2\theta$ ($5^o \leq 2\theta \leq 85^o$) with 0.0170 step size using the 'Xcelerator' check program. Infrared spectra were recorded using a Perkin-Elmer FTIR spectrophotometer employing KBr disc technique. Thermal analyses (DTA/TG) were done using the TA Instruments (UK), Model:SDTQ600, which recorded DTA and TGA simultaneously with alumina as the reference material. The experiments were carried out at a heating rate of $10^oC\ min^{-1}$ in flowing air atmosphere (flow rate:$50cm^3\ min^{-1}$). ESR measurements were carried out at room temperature using an X-band (9.839 GHz) Bruker (Germany) spectrometer with diphenyl picryl hydrazyl (DPPH) as the reference material. UV-VIS reflectance spectra were recorded using Perkin-Elmer spectrophotometer. Electron diffraction and electron transmission microscopy were carried out using FEI-Technai TEM (G-F30, Hillsboro, USA). The powder was cold-pressed into pellets of 12mm in diameter and 3mm in thickness using 3% poly vinyl alcohol (PVA) and 1% polyethylene glycol as the binders. The green pellets were then sintered at 1100$^{o}$C/2h. The densities of the sintered pellets were measured by the Archimedes principle using xylene as the liquid medium. Scanning electron microscope (SEM) (Cambridge Stereoscan S-360) was employed to study the microstructure of the sintered pellets. The capacitance measurements of the electroded pellets were carried out as a function of frequency (100Hz–1MHz) using an impedance gain-phase analyzer (HP4194A).

## 3. RESULTS AND DISCUSSION

### 3.1. Characterization of the complex oxalate precursor

#### 3.1.1. X-ray powder diffraction analysis

Fig. 1 shows the X-ray powder diffraction (XRD) pattern of the as-prepared precipitate (CCT-OX) and for the residues obtained after thermal decomposition at selected temperatures. The XRD of this complex precursor [Fig. 1.a] has been compared with those of the oxalates of individual metal ions available from International Centre for Diffraction Data (ICDD). It is clear that the pattern for the complex precursor (CCT-OX) is quite different from those of the individual oxalates. As indicated from the d-values (Table 1), there are differences in the Bragg reflections and their relative intensities. The strong lines of copper oxalate hydrate (ICDD: 00-021-0297), calcium oxalate hydrate (ICDD: 00-003-0110), and titanium oxalate hydrate ( ICDD: 00-032-1386) are different from those of the observed reflections of the as-prepared complex precursor. This confirms the appearance of the new phase as a result of the wet-chemical preparation.

#### 3.1.2. Infrared spectrum of the complex oxalate precursor

The formation of single-phase oxalate is further confirmed by the FTIR analysis. The IR spectrum of the as-prepared complex precursor is quite different from those of the individual oxalates. There are some unique features for the spectrum presented in Fig. 2(a) by way of multiplets in all the oxalate-related absorption bands. The $(C_2O_4)^{2-}$ ion can be monodentate, bidentate (chelating) or bridging (tetradentate) and rarely tridentate in metal complexes. Of these, the bidentate is the most common type. The bridging oxalate groups with the planar ($D_{2h}$) or twisted ($D_{2d}$) conformation exhibit only two absorptions arising from the C-O stretching vibrations ($\upsilon$ CO). Whereas, the bidentate oxalate groups show four $\upsilon$-type absorptions as can be expected from their $C_{2V}$ symmetry. The broad absorption band in the 1750-1620 $cm^{-1}$ region (designated as $\upsilon_1$ and $\upsilon_7$ according to the normal coordinate analyses of metal oxalate complexes by Fujita et.al [26] can be resolved into multiple peaks of which the one around 1630 $cm^{-1}$ corresponds to the bending mode of $H_2O$(hydration). The other $\upsilon_a$(C=O) absorptions are fairly overlapping that their unique features are difficult to diagnose. In contrast, the multiple absorption bands are better discernible in the region 1450-1200 $cm^{-1}$ (encompassing the absorption of $\upsilon_2$ and $\upsilon_8$ from the normal coordinate analysis). These absorptions arise from the mixed vibrations of $\upsilon_a$(CO) + $\upsilon$(C-C) as well as $\upsilon_s$(CO) + $\delta$(O-C=O). The fact that there are six or more clearly decipherable components herein indicates the presence of oxalates with both ($D_{2h}$) / ($D_{2d}$) and $C_{2V}/C_2$ point group symmetry, which is a direct evidence for the prevalence of both bridging (tetradentate) as well as the bidentate oxalate ions. The multiplets in the region 950-750 $cm^{-1}$ arise from the in-plane bending modes of mixed vibrations, $\upsilon_s$(CO)+ $\delta$(O-C=O) ($\upsilon_3$) and $\delta$(O-C=O)+ $\delta$(M-O) ($\upsilon_9$) (where M= Ti and to a lesser extent of weightage, Cu). The multiple absorption bands are also persisting in the 550-380 $cm^{-1}$ arising from the $\upsilon$(M-O) and $\delta$(O-C=O). The absorption at 540 $cm^{-1}$ will have more contribution from $\upsilon$(Ti-O) whereas the absorptions at 506 and 420 $cm^{-1}$ have dominance from $\upsilon$(Cu-O) bending. These bands are associated with mixed vibrations of $\upsilon$(C-

C) + $\delta$(O-C=O), which in turn overlap with the ring deformation absorption. The absorption bands in the lower wavenumber region of 440-380 $cm^{-1}$ are marked by $\upsilon$(M-O) which in turn correspond to $\upsilon_{10}$, $\upsilon_{11}$ or $\upsilon_5$ mixed vibrations as per the normal coordinate analysis [27].

The presence of two different types of oxalato ligands can be visualized by way of localized bonding involving the metal ions ($Ti^{4+}$,$Cu^{2+}$ and $Ca^{2+}$), wherein the interaction with $Ca^{2+}$ is envisaged to have more ionicity as compared to the other two cations [Fig.3].

### 3.1.3. Thermal analyses

In order to further verify the singularity of CCT-OX, thermal analyses (DTA/TG/DTG) were carried out on both the complex precursor as well as the individual oxalates [Figs. 4 & 5]. For quick comparison, Fig.5 (a) & (b) presents the TGA and DTA curves for the individual oxalates in air. The general features of TGA/DTA curves obtained for the as-prepared complex precursor are quite different from those of the individual oxalates. A strong exothermic peak observed around 478$^o$C and an endotherm at 753$^o$C for calcium oxalate [Fig.5.b] are altogether absent for the complex precursor [Fig.4]. Furthermore, the exothermic peak seen around 498$^o$C for the titanyl oxalate [Fig.5.b] is absent in the case of the complex precursor. This implies that the thermal decomposition behaviour of the complex precursor is uniquely different and supports the fact that the precursor synthesized is not a mixture of individual oxalates but of single–phase calcium copper titanyl oxalate.

## 3.2. Thermal decomposition of the complex oxalate precursor

Thermal decomposition of this complex precursor is studied on similar lines for alkaline earth metal titanyl oxalates reported by Murthy et.al [25,28-30]. However the decomposition behavior is found to be complex in nature due to the back reaction of CO by way of reducing Cu(II) compounds to Cu(I) or $Cu^{(O)}$ which in turn converts to CuO in air. The TGA of the as-prepared precipitate shows marked weight losses with increasing temperature in three steps. The total mass-loss observed from the TGA is around 54.8%. The theoretically expected mass-loss is ≈ 54.6%, calculated from the initial composition $CaCu_3(TiO)_4(C_2O_4)_8 \bullet 9H_2O$, which is in good agreement with the experimentally obtained value (54.8%). The possible decomposition reactions are: (i) dehydration, (ii) decomposition of the oxalate to a complex oxycarbonate and (iii) decomposition of the intermediate carbonate to calcium copper titanate.

Dehydration takes place in two steps: In the temperature range 25-130$^o$C, eight moles of water of hydration are lost. The experimentally observed weight loss of 10.7% is in reasonable agreement with the calculated value of 10.6% for the removal of 8 moles of $H_2O$.

Thermal decomposition of the oxalate takes place in 2-steps: The DTA [Fig.4] indicates only one major exothermic peak corresponding to the decomposition of oxalate. However, the careful scrutiny of the DTG curve [inset in Fig.4] reveals that the thermal decomposition takes places in two stages: The first exothermic reaction occurs in the temperature range of 180-200$^o$C. The residue at this stage has the composition: $CaCu_3(TiO)_4(C_2O_4)_5(CO_3)_3 \bullet CO_2$ as revealed by the chemical analyses of the residue after the isothermal heating at 185$^o$C for 24h and the mass-loss is ~ 6.1%. The gases evolved at this stage are carbon monoxide and $H_2O$ (vap) as per the analyses using a gas chromatograph. The residue is grey to black which on dissolution in hydrochloric acid does not leave any insolubles by way of the carbon particles arising from the disproportination of carbon monoxide.

The second step of the oxalate decomposition is the main event occurring in the thermoanalytical experiments. This exothermic reaction occurs from 230-280$^o$C which involves simultaneous evolution of carbon monoxide and carbon dioxide resulting in the intermediate carbonate as the residue. The isothermal heating at 235°C for 24 h yielded a residue with the composition: $CaCu_3Ti_4O_{11}(CO_3) \bullet CO_2$ . The mass-loss of 37.5% as compared to 38% calculated for the formation of this residue. This stage of the oxalate decomposition involves a complex set of reactions which includes decomposition of the oxalate, oxidation of CO to $CO_2$ as also the possible disproportionation of CO to $CO_2$ + C.

On further increasing the heat treatment temperature, the intermediate carbonate $CaCu_3Ti_4O_{11}(CO_3)\cdot CO_2$ decomposes to the oxycarbonate with the release of carbon dioxide retained within the matrix. This gives rise to the residue with the composition of $CaCu_3Ti_4O_{11}(CO_3)$. This occurs in the broad temperature range of 280 – 550°C.

The oxycarbonate $CaCu_3Ti_4O_{11}(CO_3)$ decomposes between 600-700°C with the evolution of carbon dioxide, giving rise to calcium copper titanate ($CaCu_3Ti_4O_{12}$). The final step in the DTG corresponds to this reaction. The observed mass-loss for the carbonate decomposition is 4.0% as against the calculated loss of 4.2%. This difference in mass-loss may be attributed to the parallel reaction as per the decomposition scheme-B (Scheme: II) that is taking place during the decomposition of the intermediate oxycarbonate; decomposition scheme-B (Scheme : II) also gives the same mass-loss as of the oxycarbonate.

The XRD pattern [Fig.1.b] of the residue $CaCu_3(TiO)_4(C_2O_4)_5(CO_3)_3 \bullet CO_2$ obtained from the isothermal heating at 185°C for 24 h does not show reflections corresponding to those of calcium carbonate, copper oxide or any one of the polymorphs of titanium dioxide (anatase/rutile/brookite). The intermediates that are formed depend upon the back reaction of the evolved gases which is influenced by the experimental conditions including the rate of heating and the accumulation of the evolved gases. The back reaction of CO by way of reducing Cu(II) compounds to Cu(I) of $Cu^{(o)}$ which in turn converts to CuO above 230°C in air. Such back reactions are most probable when the flow rate of air is less, the use of tall containers such as crucibles as against shallow boats (container geometry) and also on the quantity of the oxalate taken at a time for decomposition. If the rate of heating is slow or in the case of isothermal treatment where the sample is slowly heated from room temperature to the set temperature, the X-ray pattern [Fig.1.b] does not reveal any reflections corresponding to $CaCO_3$ , CuO or $TiO_2$. However, when the sample is introduced into the preheated furnace to >185$^oC$, the exothermic decomposition of the oxalate sets in. The XRD reflections [Fig.1.c] of low intensity observed for CuO at this stage indicate that the bulk of the residue is amorphous to X-rays.

The X-ray diffraction pattern [Fig.1.d] for the residue, $CaCu_3Ti_4O_{11}(CO_3) \bullet CO_2$ , obtained after the second stage of oxalate decomposition between 230-280$^oC$ indicates reflections (low intensity) corresponding to CuO only, whereas the reflections corresponding to $TiO_2$ and $CaCO_3$ are not detected. This indicates that, at this stage, bulk of the residue is amorphous while the minor phase of CuO formed during the thermal decomposition is crystalline to x-rays.

The X-ray diffraction pattern of the residue, $CaCu_3Ti_4O_{11}(CO_3)$, obtained at 550$^oC$ [Fig. 1.e] shows the presence of only two phases: (1) CuO and (ii) $TiO_2$ (anatase) with low intensities. The line broadening observed for CuO is due to the nanometric size of the particles and intensities are very low compared to those observed (not shown) for 100 % CuO. The reflections corresponding to $CaCO_3$ are barely discernible. This again indicates that the major portion of the residue is not crystalline to X-rays. It is evident that during the thermal decomposition of the complex oxalate precursor leading to X-ray amorphous intermediate, a parallel reaction is taking place leading to the formation of $CaCO_3$+ 3CuO+ 4$TiO_2$. Furthermore, when the precursor oxalate is thermally decomposed at 550$^oC$ in shallow alumina trays, the XRD of the residue remained totally amorphous to X-rays. It is confirmed that the phase-pure calcium copper titanate (CCTO) is formed when the complex precursor is isothermally heated above 680$^oC$ [Fig.1.f].

The infrared (i.r) spectra [Fig.2.b] of the residue, $CaCu_3(TiO)_4(C_2O_4)_5(CO_3)_3 \bullet CO_2$, from the first stage of oxalate decomposition exhibit the absorptions of both oxalate and carbonate groups. The intensities as well as the multiplicity of the oxalate absorption bands have diminished as compared to those of the as-prepared precursor. In addition, a sharp band prevails around 2340 $cm^{-1}$ which can be assigned only to the asymmetric stretching mode of carbon dioxide retained in the matrix of the X-ray amorphous residue [28]. The i.r.spectrum [Fig.2.c] of $CaCu_3Ti_4O_{11}(CO_3) \bullet CO_2$ from the second stage of oxalate decomposition shows the prevalence of ionic carbonate ($\upsilon_{asy}$ at 1509 and 1401 cm-1) as also carbon dioxide ($\upsilon_{asy}$ at 2338 $cm^{-1}$) retained within the amorphous matrix. The i.r.spectrum of the residue, $CaCu_3Ti_4O_{11}(CO_3)$, from 550$^oC$, [Fig. 2.d] shows the absorption band around 1436 $cm^{-1}$ confirming the prevalence of oxycarbonate [28] at this stage. There is no absorption band in the region 2200- 2400 $cm^{-1}$, indicating the escape of entrapped carbon dioxide. The i.r.spectrum [Fig.2.e] of the phase-pure CCTO does not exhibit absorption bands around 1436 $cm^{-1}$ corresponding to the stretching mode of the carbonate, thereby confirming the complete decomposition of the oxycarbonate at this stage. Further, there are absorption bands in the region 380-700 $cm^{-1}$ arising from the mixed vibrations of $CuO_4$ and $TiO_6$ groups prevailing in the CCTO structure. Based on the above observations, two decomposition schemes are proposed for the decomposition of CCT-OX (Scheme – II)

### 3.3. Characterization of $CaCu_3Ti_4O_{12}$ (CCTO) powders

#### 3.3.1. X-ray powder diffraction

The X-ray powder diffraction patterns of the residues obtained at the various stages of the precursor oxalate decomposition are given in Fig 1(b to f). It is confirmed that the phase-pure calcium copper titanate (CCTO) is formed when the complex precursor is isothermally heated above 680$^{o}$C [Fig 1.f]. The CCTO preparation has also been attempted by the isothermal heating of stoichiometric mixtures of individual oxalates, namely titanyl oxalate, calcium oxalate and copper oxalate in the ratio of 4:1:3. It is observed that the phase-pure CCTO could be obtained in this case only when calcinations temperature is ≥950$^{o}$C (10h) which is in agreement with the results presented in [24]. It is evident that the synthetic procedure reported therein, [24] corresponds to that of a mixture of individual oxalates and not of a single-phase oxalate precursor. The complex precursor prepared in the present work reveals the unique thermal decomposition behavior leading to the formation of phase-pure CCTO at temperatures as low as ~680$^{o}$C.

### 3.3.2. Optical reflectance spectra

It is observed that the residues remaining after the isothermal decomposition of the complex precursor $CaCu_3(TiO)_4(C_2O_4)_8 \bullet 9H_2O$, at temperatures between 150-650$^{o}$C have colors changing from green to greenish grey and to deep black. Whereas the black residue changes totally to light brown on heating above 650$^{o}$C and acquires deep yellow brown on heating at 700 to 950$^{o}$C. The brown color disappeared when the annealing temperature is increased to > 950$^{o}$C, whereupon the body color of the residues turned to steel grey. It may be reiterated that the XRD of the residue above 700$^{o}$C indicated the presence of only phase-pure CCTO. The UV-VIS reflectance spectra of the CCTO samples heat treated between 700-1000$^{o}$C have been studied because of change in the color of the residue with annealing temperature. The technique of reflectance spectroscopy is more suitable for the colored non-transparent solid samples which are insoluble in solvents.

Fig.6 shows the spectra plotted as absorbance (A = 1/R, where R is reflectance) versus the wavelength in the spectral region of 200-850nm. The CCTO samples prepared by heating the precursor at 700$^{o}$C have two broad bands: (i) around 220-550 nm and (ii) around 700 nm. The first band can be resolved by curve-fitting into 3-components with the maxima around 330, 370 and 470 nm. The intensity of these components are much higher than that of the 700nm band. The samples annealed at 900$^{o}$C exhibited lower intensity for the 700 nm band whereas the intensity of the 470nm band increased considerably. However, the spectral bands around 330 and 370 nm are discernible by curve fitting within the broad spectral envelop. The brown color of the samples can be accounted in terms of higher absorbance in the blue in comparison to the red region. For the samples heated at 1000$^{o}$C, the intensity of the band centered around 700 nm has increased considerably to the same order as that of the 220-550 nm broad band as a result of which the absorption in blue as well as red region are comparably equal, accounting for the dark body color of these samples. The intensity-maxima obtained by curve-fitting are around 330,390 and 550nm for the broad spectral band. The interpretations of the electronic spectra of copper (II) compounds are more complicated. This is because of the Jahn Teller (J-T) effect prevailing in the 3d$^{9}$ ion with the distorted octahedral structure which is counteracted by the spin-orbit coupling producing sufficient splitting of the $T_2$ ground state and thereby reducing the J-T effect less effective. Often, the J-T related vibronic interaction and the spin-orbit splitting are of comparable magnitude. As a consequence, the square coordination in Cu (II) compounds can be considered as the extreme case of Jahn-Teller distorted octahedral coordination. Thus, Cu (II) ions in the distorted octahedral, tetrahedral or square coordination exhibit complex electronic absorption spectra. Further, the strong charge-transfer band prevailing in the UV region may tail off into the blue region in the visible spectrum, rendering the Cu (II) compounds to appear brown.

The distorted tetrahedral (squashed tetrahedra) stereochemistry may account for the brown colored Cu (II) compounds. Considering the hole-formalism for the $3d^9$ ions, there are three possible transitions: (1) $d_{x^2-y^2} \leftrightarrow d_{xy}$, (2) $d_{x^2-y^2} \leftrightarrow d_{z^2}$ and (3) $d_{x^2-y^2} \leftrightarrow d_{zx}, d_{yz}$ [Fig.6]. The band around 700nm corresponds to transition-1 whereas that around 470-550nm corresponds to $d_{x^2-y^2} \leftrightarrow d_{z^2}$ (transition-2). The absorption due to $d_{x^2-y^2} \leftrightarrow d_{zx}, d_{yz}$ (transition-3) of Cu(II) maximizes around 370nm, (better discernible from the spectra of samples heated at 700$^o$C in Fig. 6) which is usually difficult to decipher in many Cu (II) complexes because of the overlap with the charge-transfer (C-T) band which tails off into the blue region. The C-T band also arises from the $O2p^6 \rightarrow Ti3d^o$ excitation of $TiO_6$ which maximizes around 330nm and is distinctly discernible in the case of the spectra of samples heated at 1000$^o$C. In CCTO, the hybridization of O(2p) and Cu(3d) orbitals are strong so that there is higher covalency for Cu-O bond than that of Ti-O so that the valence band is made up O(2p) and Cu(3d) states with less contribution from the $t_{2g}$ state of Ti($3d^o$), whereas the conduction band is mostly contributed by the Ti(3d) states. As a result of these overlapping, the absorption edge cannot be deciphered exactly.

There is a significant change in the Cu (II) electronic absorption bands when the brown colored CCTO prepared from the oxalate precursor changes to dark on heating above 950$^o$C. The energy of the $d_{x^2-y^2} \leftrightarrow d_{z^2}$ (transition-2) decreases from 2.76 to 2.42 eV (Table 2). This is accompanied by the increased intensity and hence the oscillator strength of transition-1: $d_{x^2-y^2} \leftrightarrow d_{xy}$. The energy of transition-3 also decreases, although less accurately decipherable. The curve-fitted data indicate the shift in absorption maximum from 3.35 eV for the brown sample to 3.04 eV for the dark sample heated at 1000$^o$C. These changes indicate that the copper ions are not having the same point group symmetry. The site symmetry of $D_{2h}$ for $CuO_4$ “plaquettes” is deciphered by the X-ray structure analysis of CCTO. The Cu (II) complexes are brown or yellow colored if the Cu-ions retain the four coordination with the tetrahedral geometry. The site symmetry may be somewhat lower as of $D_2$ in the brown sample which may imply the stereochemistry of nearly flattened (squashed) $CuO_4$ tetrahedra. The corresponding dipolar contribution to the effective dielectric constant of the brown colored CCTO has to be studied separately. The hybridized bonds formed between the lower 3d-states of Cu (II) ions and the oxygen 2p states are important in yielding correlated insulators as reported by Kohsaka et.al [31]. The differences in the site group symmetry of $CuO_4$ from $D_2$ to $D_{2h}$ will account for the change in color from yellow brown to black for the CCTO samples heat treated at 950 to 1000$^{o}$C. The sample heat treated at 1000$^{o}$C showed increased intensity of the band centered around 700nm which on reheating at 850$^{o}$C for 76-98h exhibited diminished intensity for the same band. This indicates that the changes in the Cu (II) geometry with heat treatment at high temperature is nearly irreversible.

### 3.3.3. Electron spin resonance spectra

The isothermal decomposition of the complex precursor $CaCu_3(TiO)_4(C_2O_4)_8 \bullet 9H_2O$, at temperatures between 150-1000$^{o}$C have different colors as mentioned earlier. It is therefore thought worthwhile to examine these residues using ESR spectroscopy to unravel the participation if any, of the point defects by way of color centres or transition metal ions of different oxidation states. The ESR spectra of the residue calcined at different temperatures were recorded at 25$^{o}$C [Fig. 7].

The ESR spectrum of the as-prepared precursor consists of a very broad band with unresolved fine structures (not presented). The hydrated precursor may have shorter relaxation times thereby leading to line-broadening. In contrast the anhydrous precursor (heated at 175$^{o}$C) gives better resolved spectra [Fig. 7.a] with partially resolvable anisotropic fine structures. The significant feature is that, the signals in Fig. 7(a) have g-values greater than that of free electrons ($g_e$= 2.0023) implying that they do not originate from the centers of higher electron density, since $\Delta g = (g_{obs} - g_e)$ is positive. For example, defect centers associated with $Ti^{3+}$ should have $g < g_e$.. Whereas, the observed spectrum can be explained in terms of hole-related centers associated with Cu (II) ($3d^9$) with S = $\frac{1}{2}$. These signals have axial symmetry in g-tensors as well as the hyperfine structures (hfs) arising from the nearly close value of nuclear magnetic moments (I) of $^{63}Cu$ and $^{65}Cu$ isotopes both having I= $\frac{3}{2}$. The ESR spectrum in Fig 7(a) arises from $Cu^{2+}$ compound of anisotropy with $g_{\perp} = 2.102$ and $g_{II} = 2.332$ and the hfs constants $A_{\perp} = 39Oe$ and $A_{II} = 173Oe$ and the line width of 314Oe. These are characteristics of tetragonally distorted copper (II) complexes involving chelated $(C_2O_4)^{2-}$ anions in the present case [32]. The intensities of the ESR signals decrease as the decomposition temperature is increased from 185-550$^o$C. Although the ESR spectrum becomes less resolvable, the g-value of the stronger signal remains unchanged. After annealing at 550$^o$C, the signals have nearly completely broadened off [Fig. 7.a]. This need not imply that the axially distorted octahedral copper has completely disappeared. The signal broadening arises from different types of interactions: spin-lattice, spin-spin, or exchange interaction as well as dipolar interaction with the likes spins. The line-broadening arising from diffusion or molecular motion is ruled out in the solid residues. The spin-spin interaction cannot be the reason for the line broadening because the spin density of the copper (II) ions remained nearly unchanged after the oxalate decomposition step. Further, the line-broadening arising from spin-lattice relaxation can also be ruled out because, the same type of oxide-related lattice prevails in these residues. Therefore the changes in dipolar interactions may be the cause for the line-broadening.

The ESR spectrum of the residue from the isothermal decomposition above 650$^o$C shows single-line symmetric signal of g=2.15 having no hyperfine structures and nearly

Lorentzian line shape [Fig. 7.b]. The difference in ESR spectra of the residue from the isothermal decomposition has to be viewed in terms of the changes in coordination around the copper (II) ions. With increasing temperatures of annealing from 650-1000$^{o}$C, the ESR line-width decreases from $\Delta H \sim$ 50-55 Oe (700$^{o}$C) to less than 40 Oe for the residue heat treated at 1000$^{o}$C. Further, the intensity of absorption increases tremendously with increasing annealing temperature [Fig.7.b].

The appearance of the single-line spectra of g = 2.15 indicates the formation of copper (II) with square planar coordination. This means that, below 650$^{o}$C, axially distorted copper (II) octahedra coexist with near square-planar copper (II), so that, the exchange interaction involving dissimilar ions takes place leading to the merger of resonance lines and signal broadening. Because of the dominance of the square planar copper (II) coordination for the residue above 650$^{o}$C, the ESR signals are discernably of single-line, with g=2.15 [Fig. 7.b]. Conversion of copper (II) from distorted octahedral to tetrahedral (squashed) and to square planar coordination takes place as the temperature of annealing is increased to ≥950$^{o}$C. The spectra of residues heated at 1000$^{o}$C are quite comparable to those reported for calcium copper titanate [33-35], which has been assigned to copper (II) in square planar coordination.

The broadening of the ESR line-width of CCTO is reported to arise from the oxygen vacancies [33,34], e.g. when annealed in argon atmosphere. The line-width decreases on reannealing in oxygen. In the present case, we observed no measurable changes in the weight of the sample when the annealing temperature is increased from 700-950$^{o}$C. The decrease in line-width and the increase in ESR intensity cannot be attributed to oxygen vacancies. Whereas, it can be better correlatively interpreted in view of the results from the optical reflectance spectra indicating the presence of squashed tetrahedral coordination of copper (II) ion. The increasing intensity in Fig. 7(b) can be explained in terms of the coexistence of squashed tetrahedra with square planar Cu (II). As the temperature of annealing is increased to ≥ 1000°C, all the Cu (II) will be acquiring square planar coordination.

**3.3.4. Transmission Electron Microscopy**

Fig.8 presents the bright field TEM images of the phase-pure CCTO powders obtained from the thermal decomposition of the oxalate precursor at 700-900°C revealing apparent microstructural features [Fig. 8 a-c ]. There are particles with curved [Fig.8 d] edges and corners. The particles are weakly agglomerated [Fig.8 d]. The size of the particles, as measured by the intercept method from the micrographs, is in the range of 30-200 nm. The TEM images [Fig.8 d ] also show the random prevalence of semi-coherent precipitate within some of the grains. These are Cu off-stoichiometric regions of CCTO as revealed by the EDS analysis with the features corresponding to exsolution lamellae.

Fig. 9(a) shows the SAED pattern with the zone axis as [012]. Spot patterns in ED indicate the single-crystallite nature of the particles. The ratio of the reciprocal vectors ($t_2/t_1$) is around 1.229 approaching the calculated value of 1.225 for the bcc lattice. The difference is marginal, and may arise from the microscope distortions. There are no elongations, streaks or distortion discernible for the diffraction spots indicating the absence of stacking faults or 2D defects. The streaks arising from the superlattice of varying dimensions along the [200] direction are reported in [14] for the sintered CCTO ceramics. It is evident that microdomain features within the grains of CCTO reported by these authors [14] are absent for the chemically prepared CCTO. Fig.9 (b) presents the HRTEM lattice image of CCTO along the (hkl) planes. The HRTEM image reveals that between the interior /exterior, there is change in the local microstructure. The slabs-width in the interiors is approx. 5.25 $\mathring{A}$ whereas in the exterior region, it is 2.68 $\mathring{A}$ . These dimensions are much lower than the unit cell parameter indicating that such features in the lattice images are not arising from supercell-subcell relation, associated with the order-disorder of cations. The (111) planes carrying the cation sublattices as the dominant electron scattering centres show polyhedral slab widths of ~ 2.7 Å as one moves to the exterior of the same crystallite [Fig.9.b]. The lattice fringes have unequal distribution of intensity. The alternative fringes retain uniform bright intensity whereas the in-between fringes have dark and bright appearance. The exterior regions with the slab–width of 2.7 Å can be demarkated which gives rise to the corrugated area of the fringes indicating that it is not arising from the variations in specimen thickness. The high resolution images [Fig. 9c] having the slab–width of 2.7 Å at the exterior regions can be due to thickness effect (Pendellosung fringes) which cannot be totally ruled out. However, by observing a large number of particles, the exterior regions [Fig. 9 c] with slab width 2.7 Å becomes thinner for powders heat treated at 1000$^o$C as compared to the powder heat treated at <950$^o$C [Fig. 9b]. This is not accompanied by the formation of oxygen vacancies as evidenced by the negligible mass loss (<0.005%) observed when the powder is heat treated at 1050$^o$C. Taking into consideration of the changes in HRTEM along with those of reflectance spectra as well as the ESR spectra for powders heat treated at two different temperature (change in the color from bright yellow to steel grey), the formation of planar defect in CCTO has been envisaged as shown in Fig.10. Herein the {hkl} type planes have specific layers carrying only (Cu+Ti), alternating with the layers carrying (Ca+Cu+Ti), oxygen being common in both the cases [Fig.10]. In the exterior of the crystallites, if the $CuO_4$ groups are of squashed tetrahedral rather than square planar geometry, the (Cu+Ti) layers are pushed apart from the layers carrying (Cu+Ca+Ti). The distortion of the (Cu+Ti) containing layers will be more conspicuous than that of the adjoining (Ca+Cu+Ti) layers because of the presence of larger sized $Ca^{2+}$ with higher (dodecahedral) coordination and ionicity. The deviation from the square planar geometry of copper will influence all the (Cu+Ti) containing layers since each $TiO_6$ is corner-sharing oxygen with six $CuO_4$ groups. The latter, in turn, are isolated from one another and the planar $CuO_4$ are oriented in pairs along all the three crystallographic directions. This will lead to the lattice fringes with the slab–width of ~ 2.7 Å at the exterior regions in the crystallite. The inter-conversion between squashed tetrahedral and square planar plaquettes of $CuO_4$ will also influence the stereochemical disposition of $TiO_6$ octahedra. The corresponding changes in crystal structure of the *bcc* titanate, if any, could not be detected by XRD. The structurally continuous inter-conversion between perovskite and brownmillerite can be brought out for rough comparison with the present case, wherein the involvement of oxygen vacancies is inevitable. Presently, the role of oxygen non-stoichiometry is limited and may be prevailing at temperatures >1050$^oC$. Thus, the microstuctural studies show that the yellow-brown color of CCTO crystallites arises from the $CuO_4$ with the squashed tetrahedral geometry prevailing in the exterior regions of the chemically prepared powder specimens.

### 3.3.5. Dielectric characteristics of CCTO ceramics

Figure.11 shows the frequency (100Hz-100 kHz) dependence of room-temperature dielectric constants of CCTO disks sintered at 1100$^oC$(2h) processed from powders obtained from the oxalate precursor. These disks have improved density of >95% eventhough the duration of sintering has been shorter (≤ 2h). This is indicative of the improved reactivity of the powders having high sinterability than that of the ceramically prepared powders. The microstructural studies by SEM showed that the grain sizes are in the range of 40-100 μm with no discernible grain boundary phases. There is anomalously high dielectric constant for these ceramic samples, which is in agreement with the previous publications. The dielectric constant decreases from 43,000 to 32,700 while the dielectric loss increases from 0.065 to 0.12 as the frequency increases from 100Hz to 100 kHz. The temperature dependence of dielectric constant measured at 1 kHz indicated that it is nearly constant in the temperature range of 100-600K. Although widely different explanations have been proposed for the origin of high dielectric constants [3-17], the issue remains unsettled with the lack of direct evidences.

## 4. CONCLUSIONS

A wet chemical method has been developed for the preparation of complex oxalate precursor, $CaCu_3(TiO)_4(C_2O_4)_8 \bullet 9H_2O$. The precursor gives rise to nano-crystalline phase pure $CaCu_3Ti_4O_{12}$ (CCTO) powders with a crystallite size varying from 30-200nm when heat treated at > 680$^o$C. The evolution of CCTO phase with increasing temperature of annealing is associated with the changing stereochemistry of Cu(II) ions from distorted octahedral to squashed tetrahedral and further to square planar co-ordination. The significant observation during the present studies is the coexistence of nearly flattened tetrahedral Cu (II) with those of square planar co-ordination. The microstructural features as revealed by the HRTEM indicate that $CuO_4$ groups with the squashed tetrahedral geometry is more prevalent in the exterior regions of the powder particles as compared to the predominance of square planar $CuO_4$ in the interior. The giant dielectric constant of CCTO can be envisaged to arise from the dipolar contribution of tetrahedral $CuO_4$ with the relaxational fluctuations in specific regions within the crystallites. The high dielectric constant for CuO ceramics has been reported [36] recently, which brings forth the fact that the $CuO_4$ related dielectric contribution in CCTO can be substantially significant.

## Acknowledgements

The management of Central Power Research Institute are acknowledged for the financial support (CPRI Project No.5.4.49).

## References

1. A. Deschanvres, B. Raveau, and F. Tollemer, Substitution of Copper for a divalent metal in perovskite-type titanates , Bull. Soc. Chim. Fr. (1967) 4077-4078.

2. B. Bochu, M. N. Deschizeaux, J. C. Joubert, A.Collomb, J.Chenavas and M.Marezio, Synthesis and characterization of a series of perovskite titanates isostructural with $CaCu_3](Mn_4)O_{12}$ J. Solid State Chem. 29 (1979) 291-298.

3. M. A. Subramanian,D.Li.N.Duran,B.A.Reisner,A.W.Sleight, High Dielectric Constant in ACu3Ti4O12 and ACu3Ti3FeO12 Phases, J. Solid State Chem. 151 (2000) 323-325.

4. C.C.Homes,T.Vogt,S.M.Shapiro,S.Wakimoto,A.P.Ramirez, Optical response of high-dielectric-constant perovskite-related oxide, Science. 293 (2001) 673-676.

5. T.B. Adams, D.C. Sinclair, and A.R.West, Giant barrier capacitance effects in $CaCu_3Ti_4O_{12}$ ceramics, Adv. Mater. 14 (2002) 1321-1323.

6. B.A. Bender, M.-J. Pan, The effect of processing on the giant dielectric properties of $CaCu_3Ti_4O_{12}$ , Mater.Sci. Eng., B. 117 (2005) 339-347.

7. L. Ni, X.M. Chen , X.Q. Liu, R.Z. Hou, Microstucture-dependent giant dielectric response in $CaCu_3Ti_4O_{12}$ ceramics , Solid State Commun. 139 (2006) 45-50.

8. D.C. Sinclair, T.B. Adams, F.D. Morrison, A.R. West, $CaCu_3Ti_4O_{12}$ : One-step internal barrier layer capacitor, Appl. Phys. Lett. 80 (2002) 2153-2155.

9. J.Liu, C. Duan, W.G. Yin, W.N. Mei, R.W. Smith, J.R.Hardy, Large dielectric constant and Maxwell-Wagner relaxation in $Bi_{2/3}Cu_3Ti_4O_{12}$ , Phys.Rev. B. 70 (2004)144106-1 – 144106-6.

10. J.Liu, C. Duan, W.N. Mei, R.W. Smith, J.R.Hardy, Dielectric properties and Maxwell-Wagner relaxation of compounds $ACu_3Ti_4O_{12}$ (A=Ca,$Bi_{2/3}$,$Y_{2/3}$,$La_{2/3}$) J.Appl. Phy. 98 (2005) 93703-1- 93703-5.

11. P. Lunkenheimer, V. Bobnar, A.V. Pronin, A.I. Ritus, A.A. Volkov, and A. Loidl, Origin of apparent colossal dielectric constants, Phys. Rev. B. 66 (2002) 52105-1 - 52105-4.

12. P. Lunkenheimer, R. Fichtl, S.G. Ebbinbhaus, A.Loidl, Nonintrinsic origin of the colossal dielectric constants in $CaCu_3Ti_4O_{12}$ , Phys.Rev.B. 70 (2004) 172102-1 -172102-4.

13. S.Y.Chung, Lattice distortion and polarization switching in calcium copper titanate, Appl. Phys. Lett.87 (2005) 52901.

14. T.T. Fang and C. P. Liu, Evidence of the Internal Domains for Inducing the Anomalously High Dielectric Constant of $CaCu_3Ti_4O_{12}$ , Chem. Mater. 17 (2005) 5167-5171.

15. J. Li, K. Cho, N. Wu, and A. Ignatiev, Correlation Between Dielectric Properties and Sintering Temperatures of Polycrystalline $CaCu_3Ti_4O_{12}$, IEEE Trans. Dielectr. Electr. Insul. 11 (2004) 534-541.

16. L.Fang and M. Shen, Effects of postanneal conditions on the dielectric properties of CaCu3Ti4O12 thin films prepared on Pt/Ti/$SiO_2$ /Si substrates, J.Appl. Phys. 95 (2004) 6483-6485.

17. T.T.Fang, L.T. Mei and H.F. Ho, Effects of Cu stoichiometry on the microstructures, barrier-layer structures, electrical conduction, dielectric responses,and stability of $CaCu_3Ti_4O_{12}$, Acta Mater. 54 (2006) 2867-75.

18. Y.Zhu,J.C.Zheng, L.Wu,A.I.Frenkel,J.Hanson,P.Northrup and W.Ku, Nanoscale disorder in $CaCu_3Ti_4O_{12}$ : A New route to the enhanced dielectric response, Phys.Rev.Lett, 99 (2007) 037602-1 – 037602-4.

19. L.Wu, Y.Zhu, S.Park, S.Shapiro ,G.Shirane and J.Tafto, Defect structure of the high-dielectric-constant perovskite $CaCu_3Ti_4O_{12}$, Phys.Rev B, 71 (2005) 014118-1 – 014118-7.

20. S.Jin, H.Xia, Y.Zhang, J. Guo and J. Xu, Synthesis of $CaCu_3Ti_4O_{12}$ ceramic via a sol-gel method , Mater.Lett. 61 (2007) 1404-1407.

21. J.Liu, Y.Sui, C.Duan, W.Mei, R.W. Smith, and J.R. Hardy, $CaCu_3Ti_4O_{12}$ : Low-temperature synthesis of pyrolysis of an organic solution, Chem. Mater. 18 (2006) 3878.

22. P. Jha, P. Arora, A.K. Ganguli, Polymeric citrate precursor route to the synthesis of the high dielectric constant oxide, $CaCu_3Ti_4O_{12}$ , Mater.Lett. 57 (2003) 2443.

23. W.Ren, Z.Yu, V.D.Krstic and B.K.Mukherjee, Structure and properties of high dielectric constant $CaCu_3Ti_4O_{12}$ ceramics, IEEE Intl. Ultrasonic, Ferroelectrics and Frequency Control Joint 50th Anneversary Conference, Montreal, Canada, 24-27, August 2004, p149.

24. S. Guillemet-Fritsch, T. Lebey , M. Boulos and B. Durand, Dielectric properties of $CaCu_3Ti_4O_{12}$ based multiphased ceramics, J. European Ceram.Soc, 26 (2006) 1245-1257.

25. H.S.G. Murthy, M.S. Rao and T.R.N. Kutty, Thermal decomposition of titanyl oxalates-IV, Strontium and calcium titanyl oxalates , Thermochim.Acta. 13 (1975) 183-91.

26. J.Fujita, A.E. Martell, and K.KNakamoto, Infrared Spectra of Metal Chelate Compounds. VI. A Normal Coordinate Treatment of Oxalato Metal Complexes , J.Chem.Phys. 36 (1962) 324-331 & Infrared Spectra of Metal Chelate Compounds. VII. Normal Coordinate Treatments on 1:2 and 1:3 Oxalato Complexes , J.Chem.Phys.36 (1962) 331-338.

27. K. Nakamoto, Infrared and Raman Spectra of Inorganic and Coordination Compounds, John WiIey & Sons, Jew York, 3rd edn. (1978) part III, pp.233-237.

28. H.S.G. Murthy, M.S. Rao and T.R.N. Kutty, Thermal decomposition of titanyl oxalates-I, Barium titanyl oxalate, J.Inorg.Nucl.Chem. 37 (1975) 891-898.

29. H.S.G. Murthy, M.S. Rao and T.R.N. Kutty, Thermal decomposition of titanyl oxalates-II, Kinetics of decomposition of barium titanyl oxalate, J.Inorg.Nucl.Chem. 37 (1975) 1875-1878.

30. H.S.G. Murthy, M.S. Rao and T.R.N. Kutty, Thermal decomposition of titanyl oxalates-III, Lead titanyl oxalate, J.Inorg.Nucl.Chem. 38 (1976) 417-419.

31. Y. Kohsaka, C. Taylor, K. Fujita, A. Schmidt, C. Lupien, T. Hanaguri, M. Azuma, M. Takano, H. Eisaki, H. Takagi, S. Uchida, and J. C. Davis, An Intrinsic Bond-Centered Electronic Glass with Unidirectional Domains in Underdoped Cuprates, Science. 315 (2007) 1380 -1385.

32. D.Kivelson and R.Neiman, ESR Studies on the Bonding in Copper Complexes, J.Chem.Phys. 35 (1961) 149-155.

33. M. A. Pires, C. Israel, W. Iwamoto, R. R. Urbano, O. Agüero, I. Torriani, C. Rettori, P. G. Pagliuso, L. Walmsley, Z. Le , J. L. Cohn and S. B. Oseroff , Role of oxygen vacancies in the magnetic and dielectric properties of the high-dielectric-constantsystem CaCu3Ti4O12: An electron-spin resonance study, Phys. Rev. B. 73 (2006) 224404-1-224404-7.

34. M. C. Mozzati, C. B. Azzoni, D. Capsoni, M. Bini, and V. Massarotti, Electron paramagnetic resonance investigation of polycrystalline $CaCu_3Ti_4O_{12}$, J. Phys.: Condens. Matter. 15 (2003) 7365-7374.

35. D. Capsonia, M. Binia, V. Massarottia, G. Chiodellib, M.C. Mozzatica, C.B. Azzonic, Role of doping and CuOsegregation in improving the giant permittivity of $CaCu_3Ti_4O_{12}$ , J. Solid State Chem. 177 (2004) 4494–4500.

36. S. Sarkar, P.K. Jana, B. K. Chaudhuri, and H. Sakata, Copper (II) oxide as a giant dielectric material, Appl. Phys. Lett. 89 (2006) 212905.

## Figure Captions

**Figure. 1**. X-ray diffraction patterns of complex precursor (a) as prepared (b) slow heating at 185$^{o}$C, (c) fast heating at 185$^{o}$C, (d) heat treated at 235$^{o}$C, (e) heat treated at 550$^{o}$C , (f) heat treated at 700$^{o}$C ( phase-pure CCTO) and (g) CCTO from the ICDD data file card no. 01-075-1149

**Figure. 2** FTIR spectra of complex precursors (a) as prepared precursor, (b) at 185$^{o}$C, (c) at 235$^{o}$C , (d) at 550$^{o}$C and (e) for phase-pure CCTO.

**Figure. 3.** Local bonding in the multimetal-oxalato complex of CCT-OX accounting for bridging as well as bidentate ligands.

**Figure. 4**. Simultaneous DTA/TGA for complex precursor CCT-OX at a heating rate of 10$^{o}$C min$^{-1}$ in air atmosphere (flow 50cm$^{3}$ min$^{-1}$). The inset shows the DTG recorded for the complex precursor in the range of 220-325$^{o}$C.

**Figure .5**. (a) TGA curves of the individual oxalates namely, calcium oxalate, titanyl oxalate and copper oxalate, (b) DTA of the individual oxalates, calcium oxalate, titanyl oxalate and copper oxalate.

**Figure. 6**. Reflectance spectra of CCTO heat treated at different temperatures, Absorbance, (A=1/R, where R=reflectance) , is plotted on y-axis.

**Figure. 7**. ESR spectra recorded at 25$^{o}$C for the complex precursor heat treated at different temperatures between (a) 185-550$^{o}$C and (b) 650-1000$^{o}$C. (DPPH= diphenyl picryl hydrazyl )

**Figure 8** . Bright field TEM images of phase -pure CCTO obtained at 680$^{o}$C (a-c) particle dimensions ranging from 30-200nm (d) weakly agglomerated particles (curved edges and corners) having the coherent precipitates within the grains retaining exsolution features (The background circles are features from the holey carbon grid used ).

**Figure 9**. (a) SAED pattern obtained (annealed at 900$^{0}$C/8h) with the zone axis of [012], $t_2/t_1 = 1.229$ , (b) HRTEM lattice image of CCTO along the (hkl) planes carrying the cation sublattices as the dominant electron scattering centres showing polyhedral slab widths of 2.7 Å in the exterior regions of the same crystallite (demarcated for easy visual comparison) and (c) HRTEM image of CCTO heat treated at 1000$^{o}$C showing that the lattice fringes ~2.7 Å becomes thinner at the exterior regions.

**Figure. 10**. Showing the alignment of the Ca,Cu ,Ti and O positions in the cubic cell of CCTO account for polyhedral slab widths in the shell regions as the coordination polyhedra moves apart in the exterior parts of the same crystallite. (The size of oxygen is purposely shown smaller)

**Figure. 11**. Frequency dependence of room-temperature dielectric properties for the disk sintered at 1100$^{o}$C (2h).

## Scheme Captions

Scheme : I. The different schemes attempted for the preparation of the complex oxalate precursor.

Scheme: II. Reactions occurring during the thermal decomposition of the complex oxalate precursor : Schemes A (major process) and Scheme B (minor process ).

## Table Captions

**Table.1** . X-ray diffraction data on the as prepared complex precursor, $CaCu_3(TiO)_4(C_2O_4)_8 \bullet 9H_2O$.

**Table.2.** Energy (eV), ratio of the intensities for the various transitions.

Table 1. X-ray diffraction data on the as prepared complex precursor, $CaCu_3(TiO)_4(C_2O_4)_8 \bullet 9H_2O$.

| Pos. [°2 θ] (deg) | d-spacing [Å] | Rel. Int. [%] |
|---|---|---|
| 8.528 | 10.368 | 13 |
| 9.261 | 9.550 | 11 |
| 9.864 | 8.967 | 12 |
| 13.269 | 6.672 | 2 |
| 18.241 | 4.864 | 3 |
| 22.846 | 3.893 | 100 |
| 28.713 | 3.109 | 3 |
| 36.001 | 2.495 | 8 |
| 36.976 | 2.431 | 6 |
| 38.902 | 2.315 | 8 |
| 42.189 | 2.142 | 4 |
| 46.560 | 1.951 | 4 |
| 51.629 | 1.769 | 7 |

Table. 2 Energy (eV), ratio of the intensities for the various transitions.

| Sample | Energy (in eV ) for the different transitions | | | Ratio of the intensities ($I_1/I_2$) |
|---|---|---|---|---|
| | Transition 1 | Transition 2 | Transition 3 | |
| 700$^oC$ | 1.77 | 2.76 | 3.35 | 0.66 |
| 900$^oC$ | 1.81 | 2.69 | 3.29 | 0.53 |
| 1000$^oC$ | 1.78 | 2.42 | 3.04 | 0.95 |

**Figure.1**

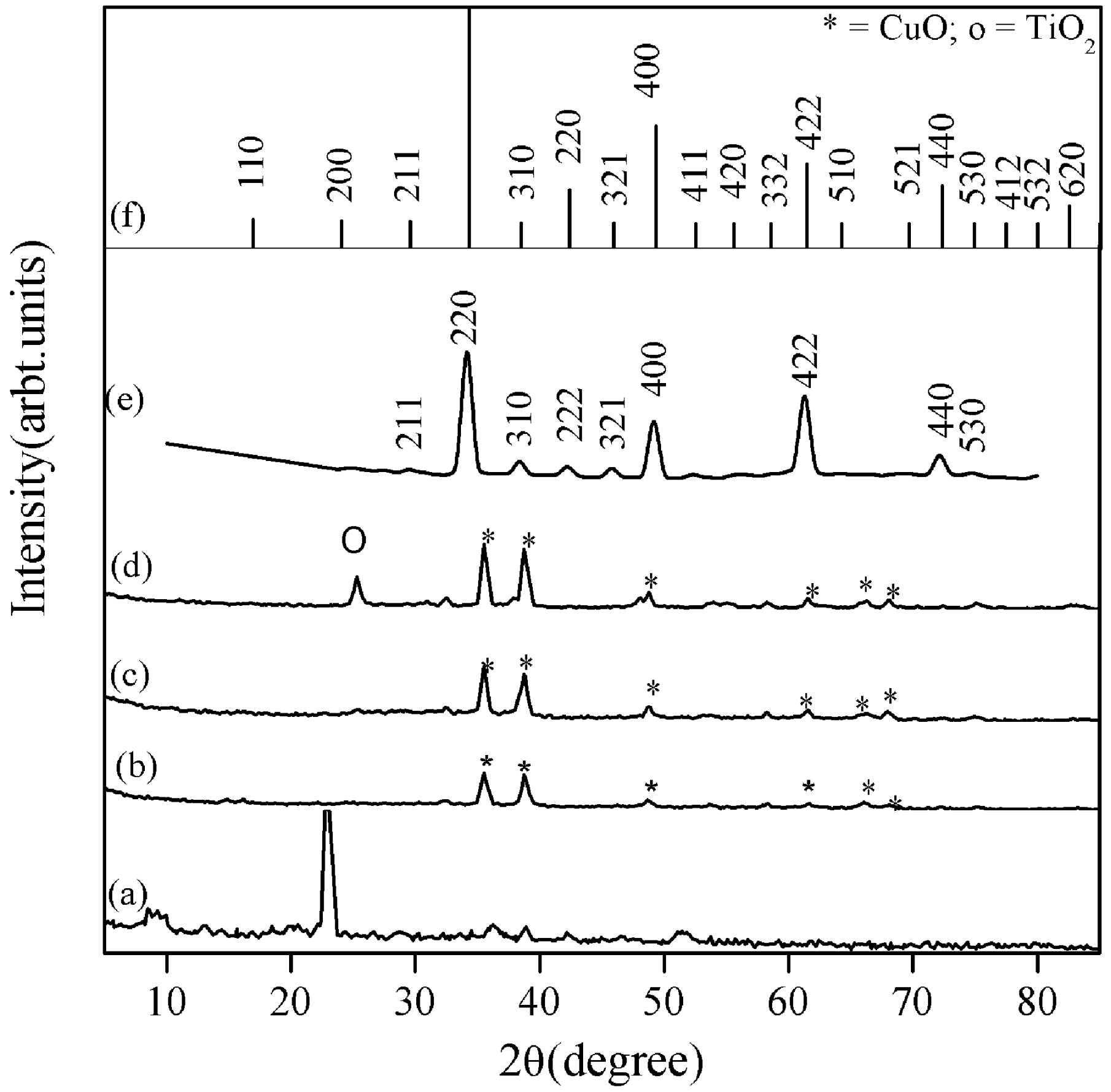

**Figure.2**

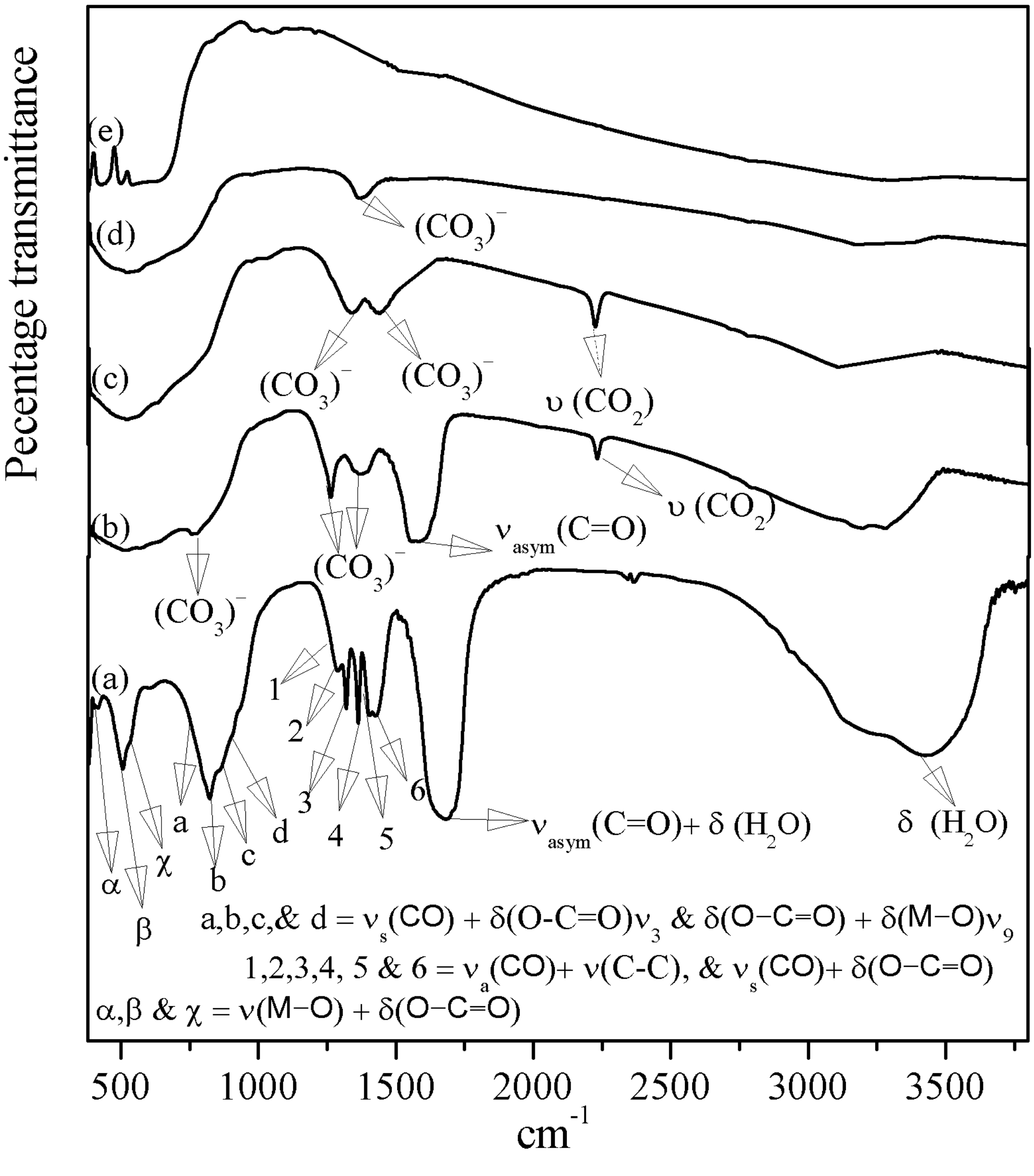

Pecentage transmittance
(e)
(d)
(c)
(b)
(a)
(CO3)⁻
(CO3)⁻
(CO3)⁻
υ (CO2)
υ (CO2)
νasym(C=O)
(CO3)⁻
(CO3)⁻
1
2
3
4
5
6
a
b
c
d
α
β
χ
νasym(C=O)+ δ (H2O)
δ (H2O)
a,b,c,& d = νs(CO) + δ(O-C=O)ν3 & δ(O−C=O) + δ(M−O)ν9
1,2,3,4, 5 & 6 = νa(CO)+ ν(C-C), & νs(CO)+ δ(O−C=O)
α,β & χ = ν(M−O) + δ(O−C=O)
500
1000
1500
2000
2500
3000
3500
cm-1

Figure.3

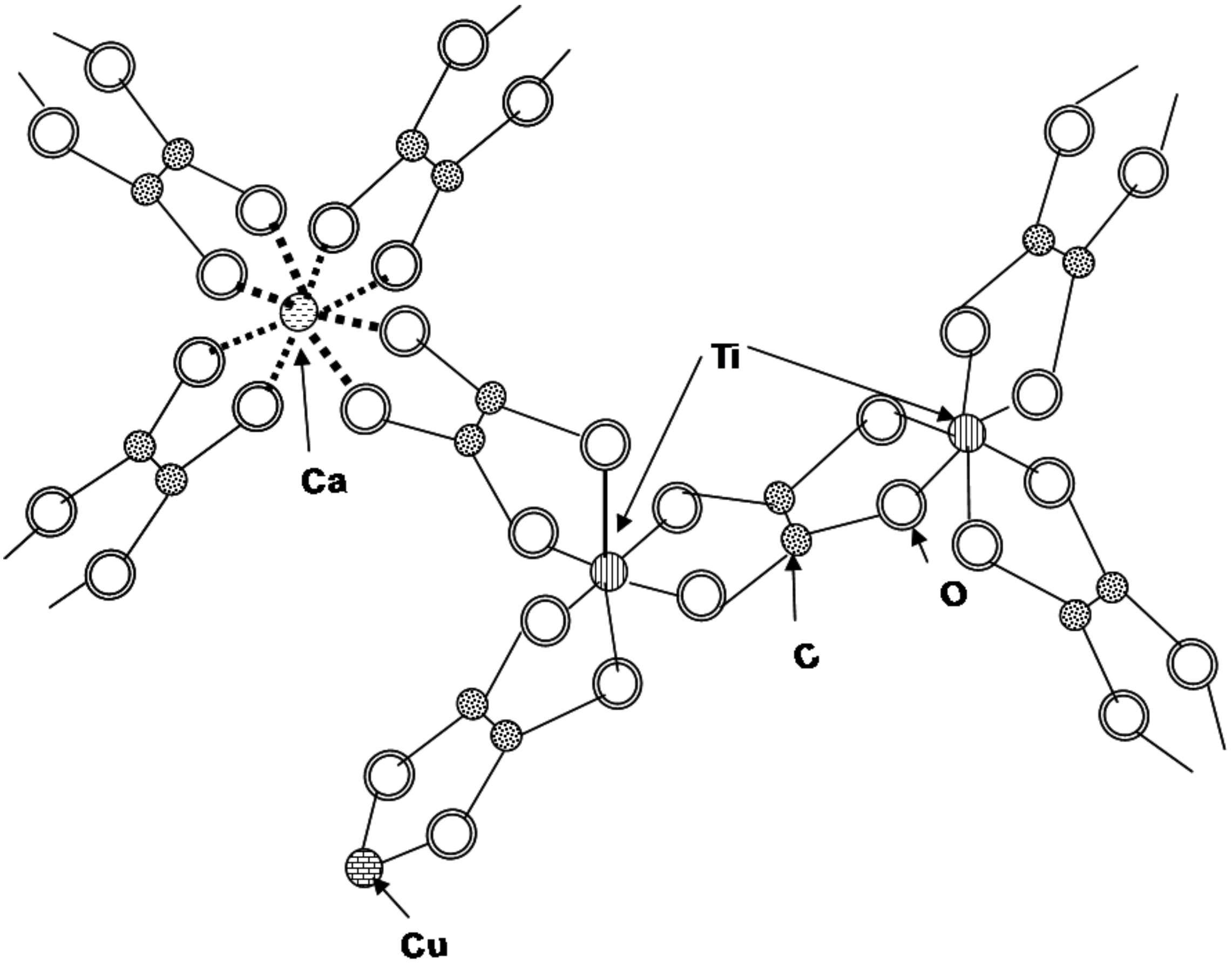

**Figure.4**

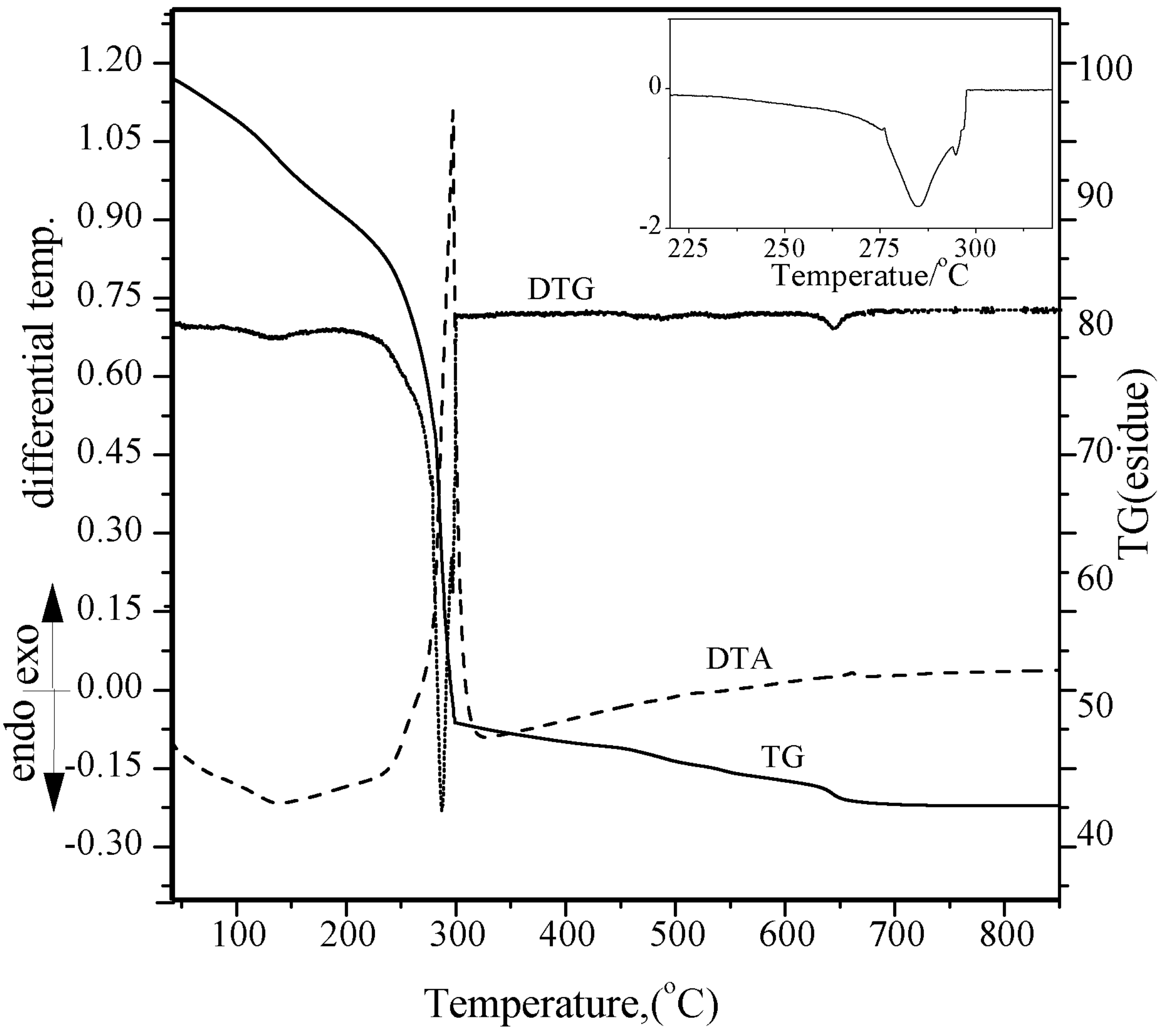

**Figure.5**

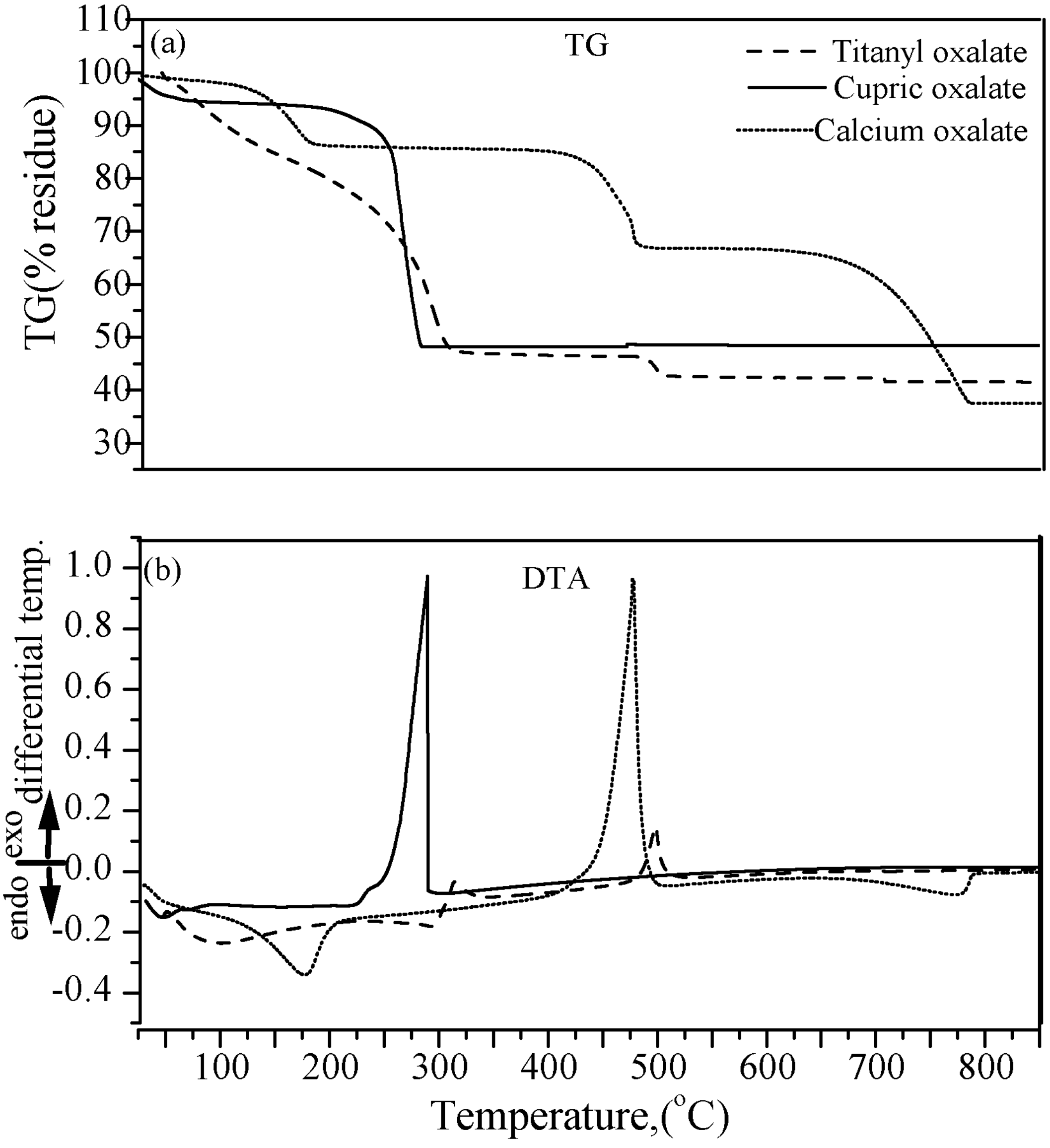

**Figure.6**

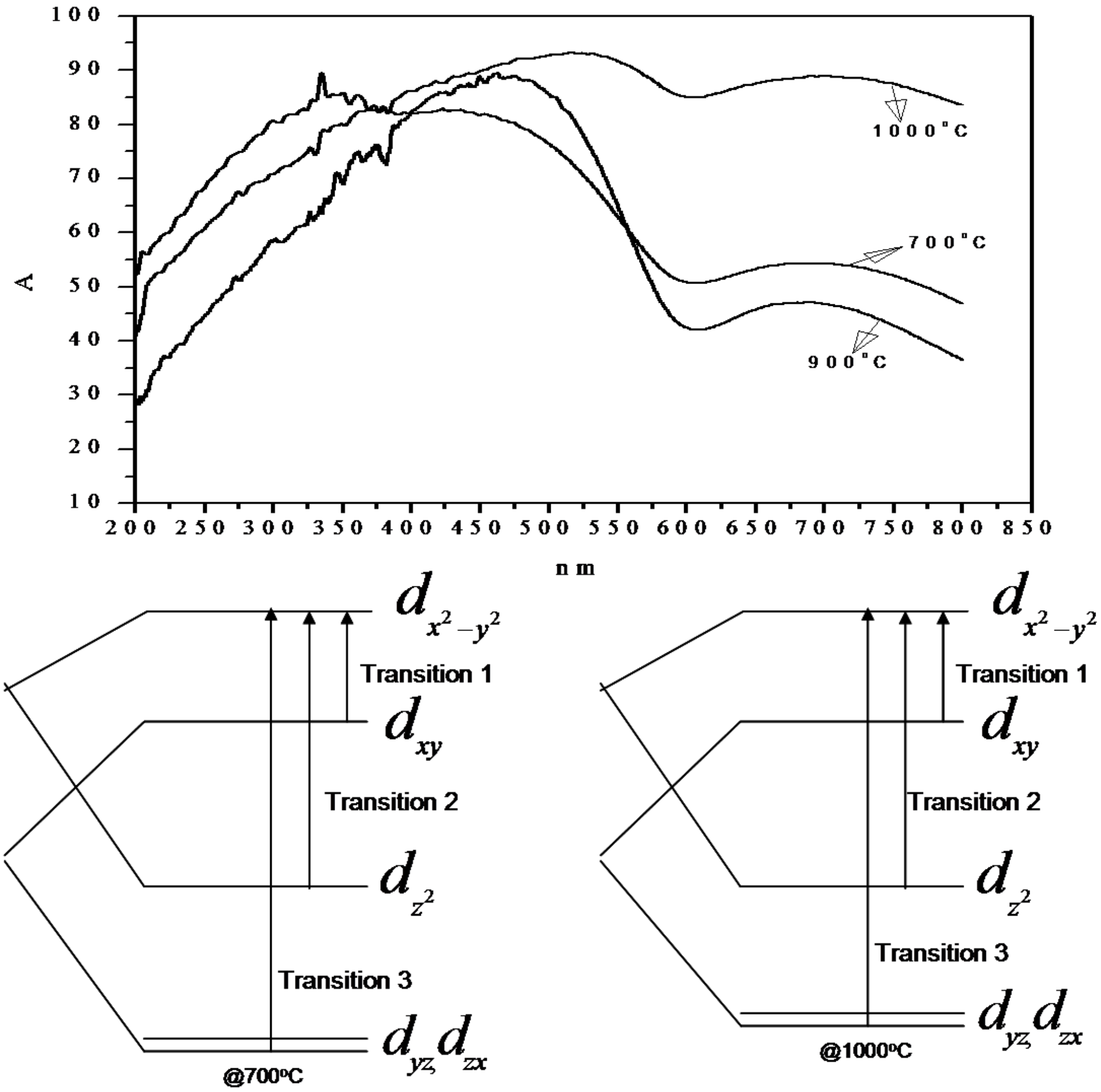
100
90
80
70
60
50
40
30
20
10
A
200 250 300 350 400 450 500 550 600 650 700 750 800 850
nm
1000°C
700°C
900°C
Transition 1
Transition 2
Transition 3
@700ºC
@1000ºC
$d_{x^2-y^2}$
$d_{xy}$
$d_{z^2}$
$d_{yz}, d_{zx}$

**Figure.7(a)**

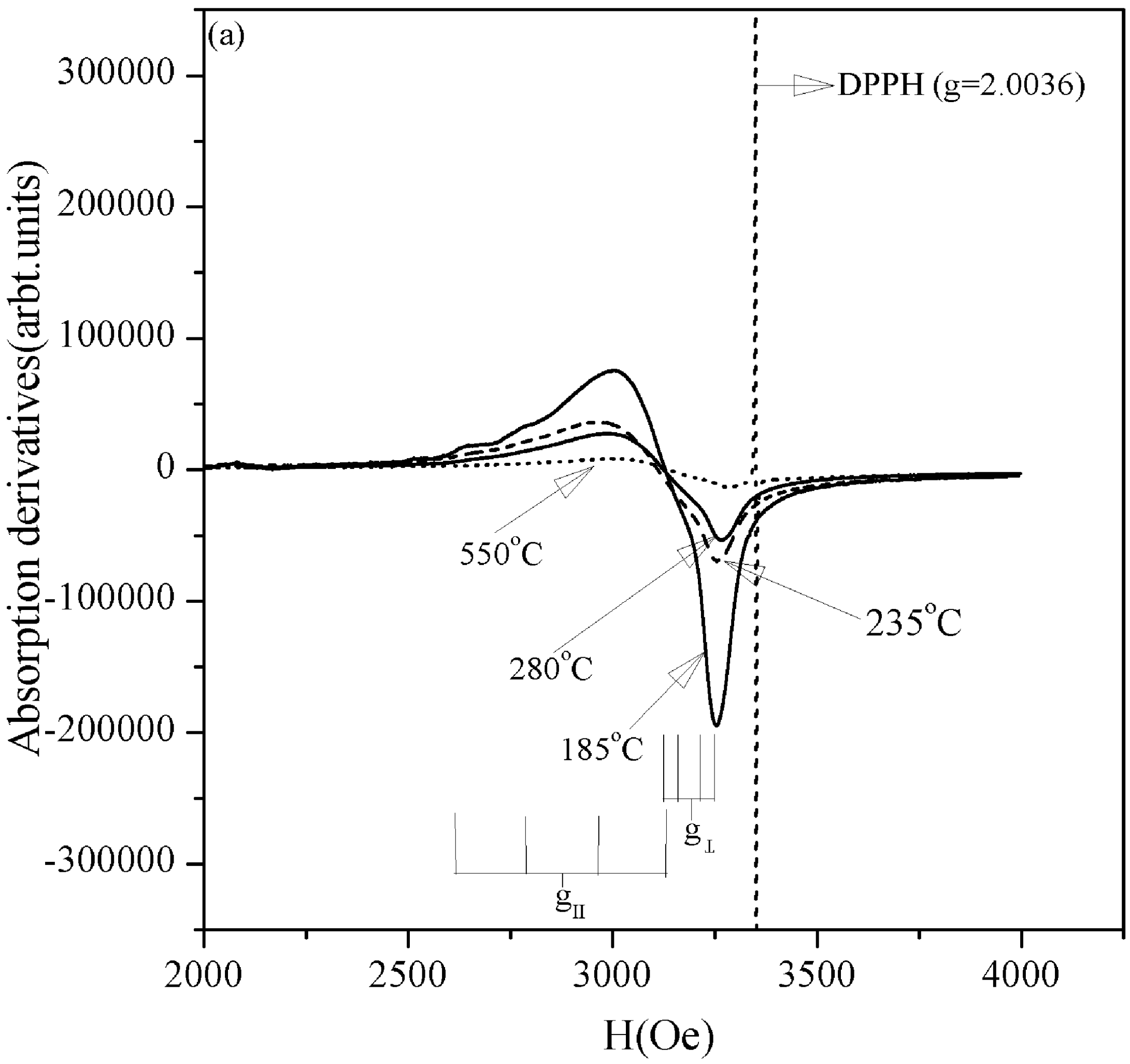

**Figure.7(b).**

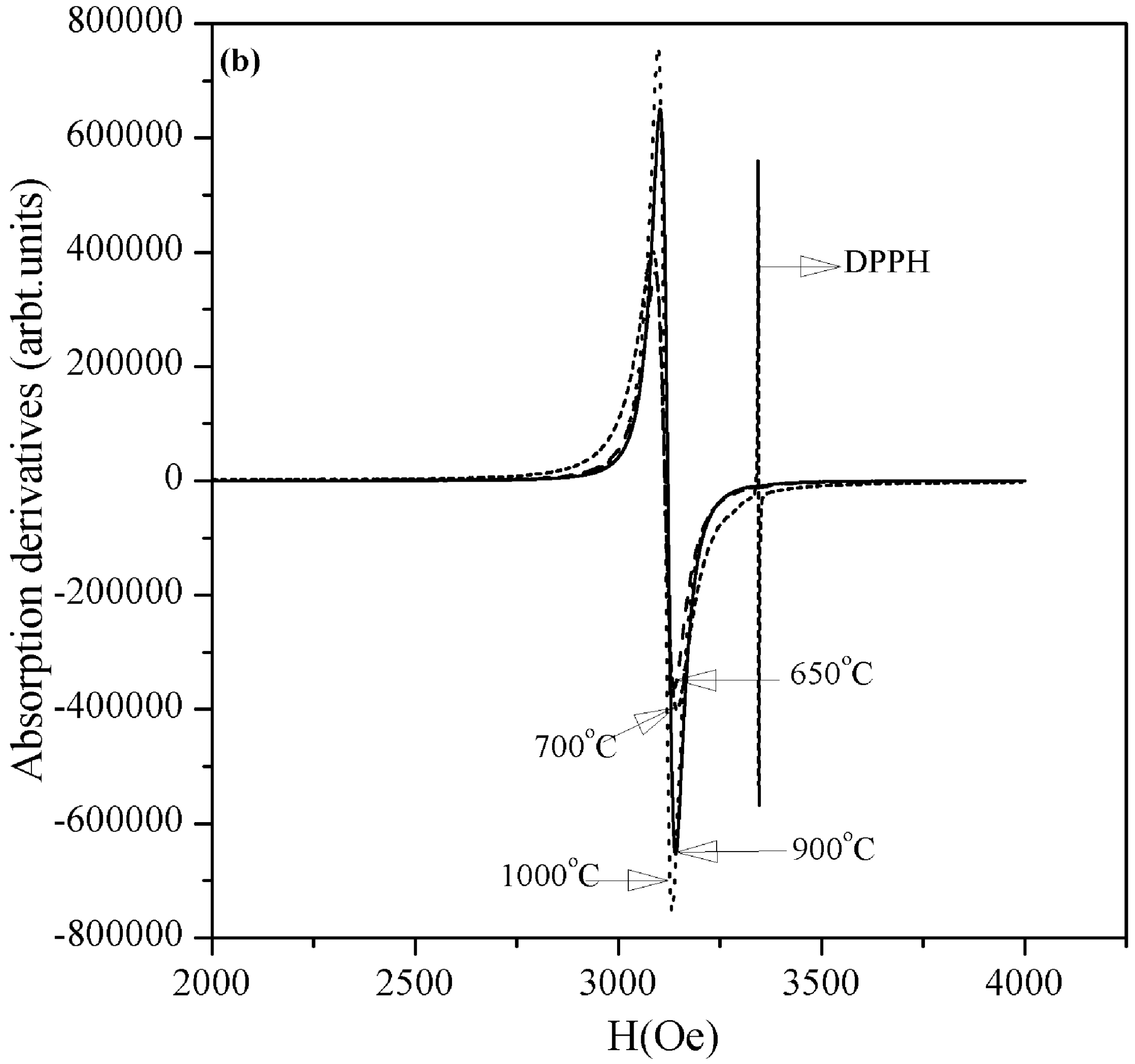

**Figure.8 (a-d)**

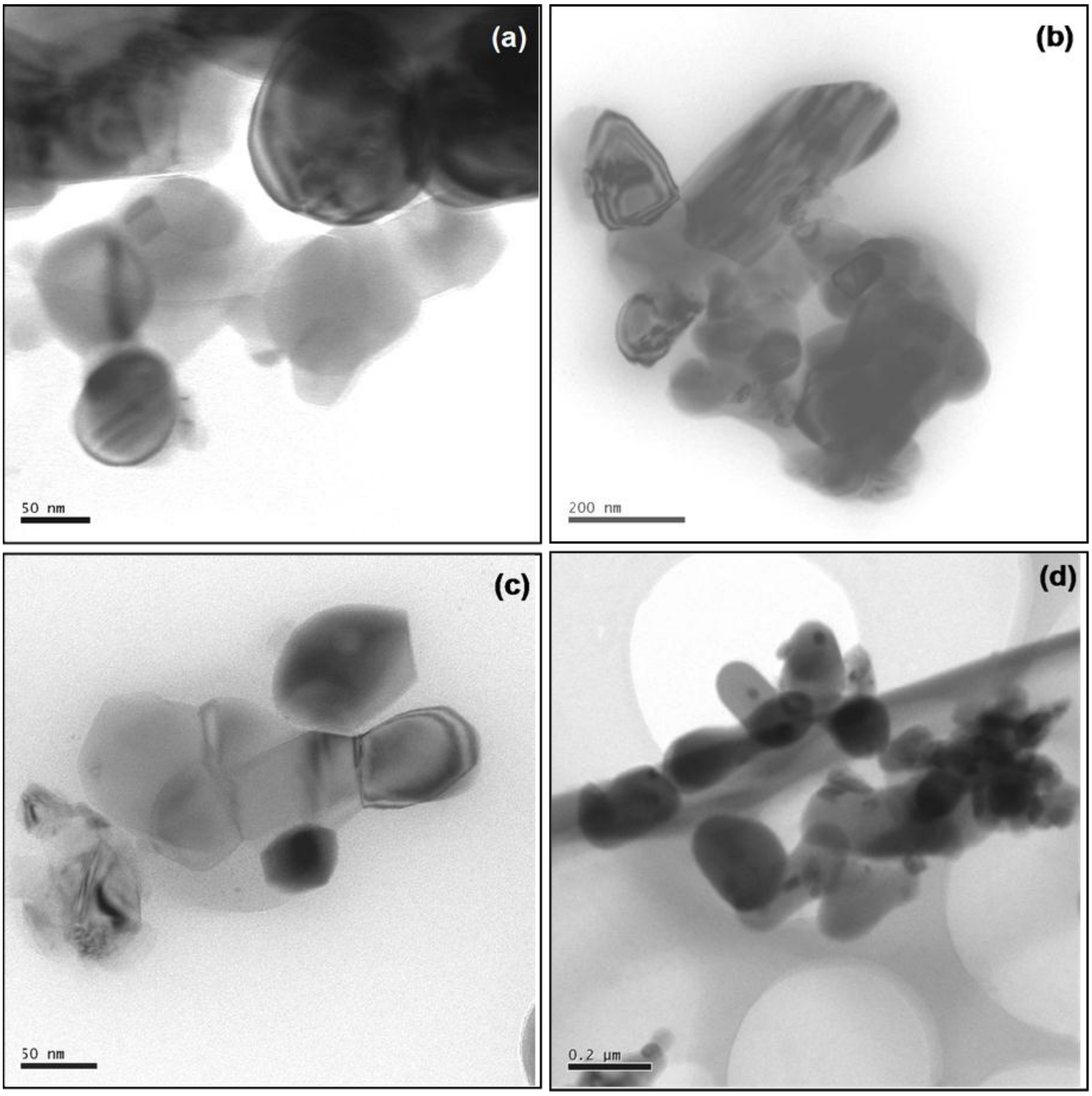

**Figure.9(a-c)**

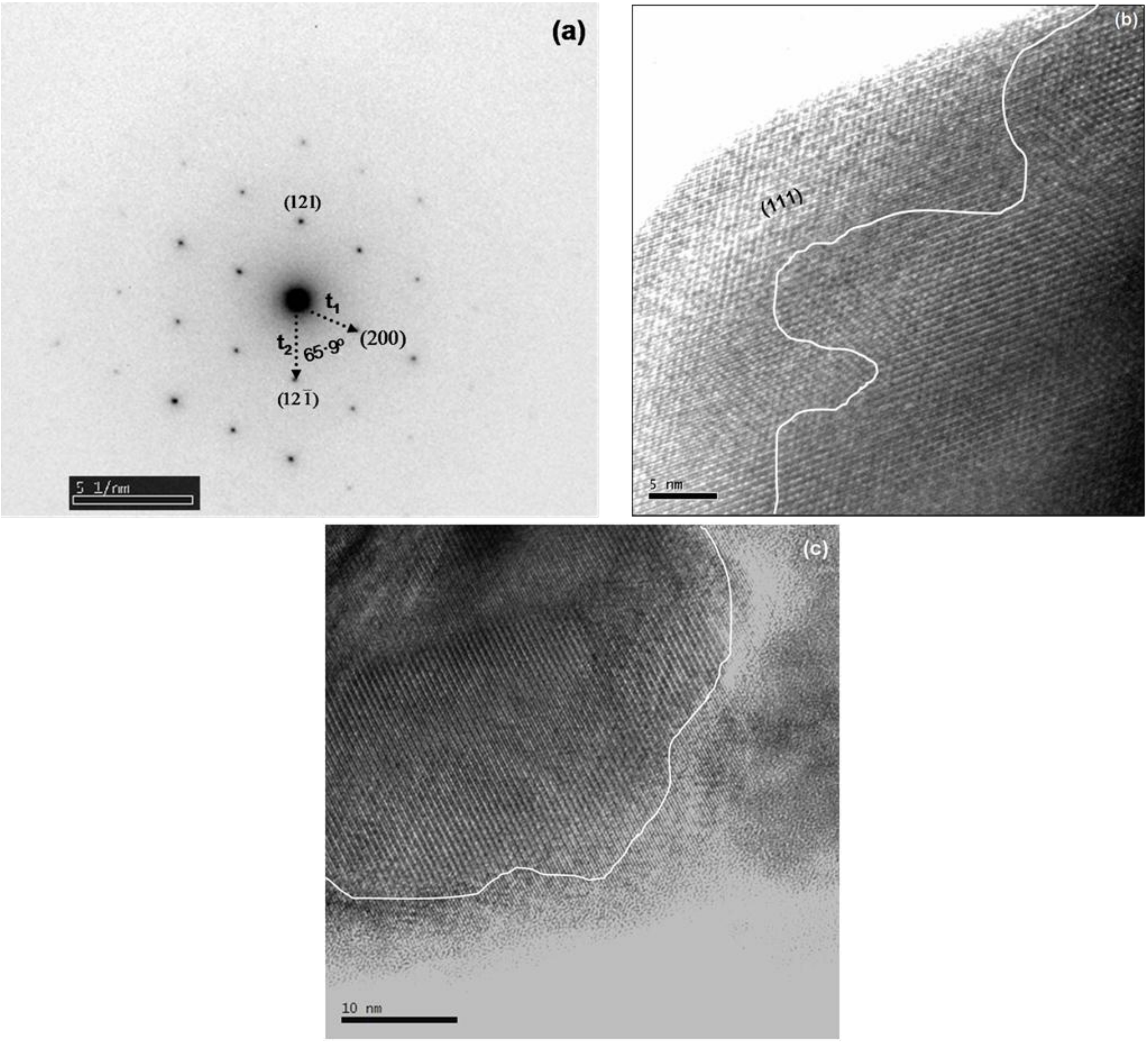

**Figure.10**

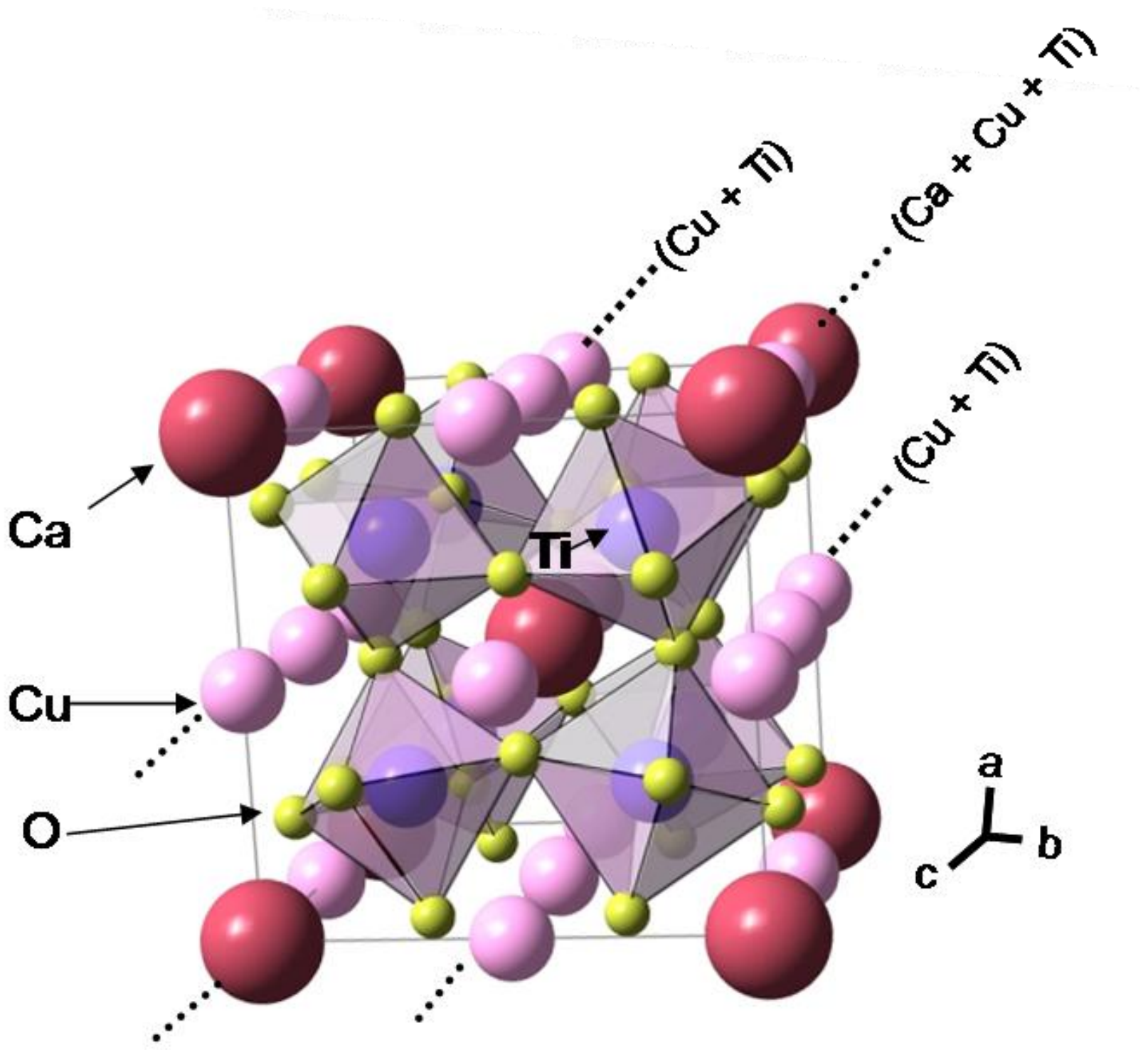

**Figure. 11**

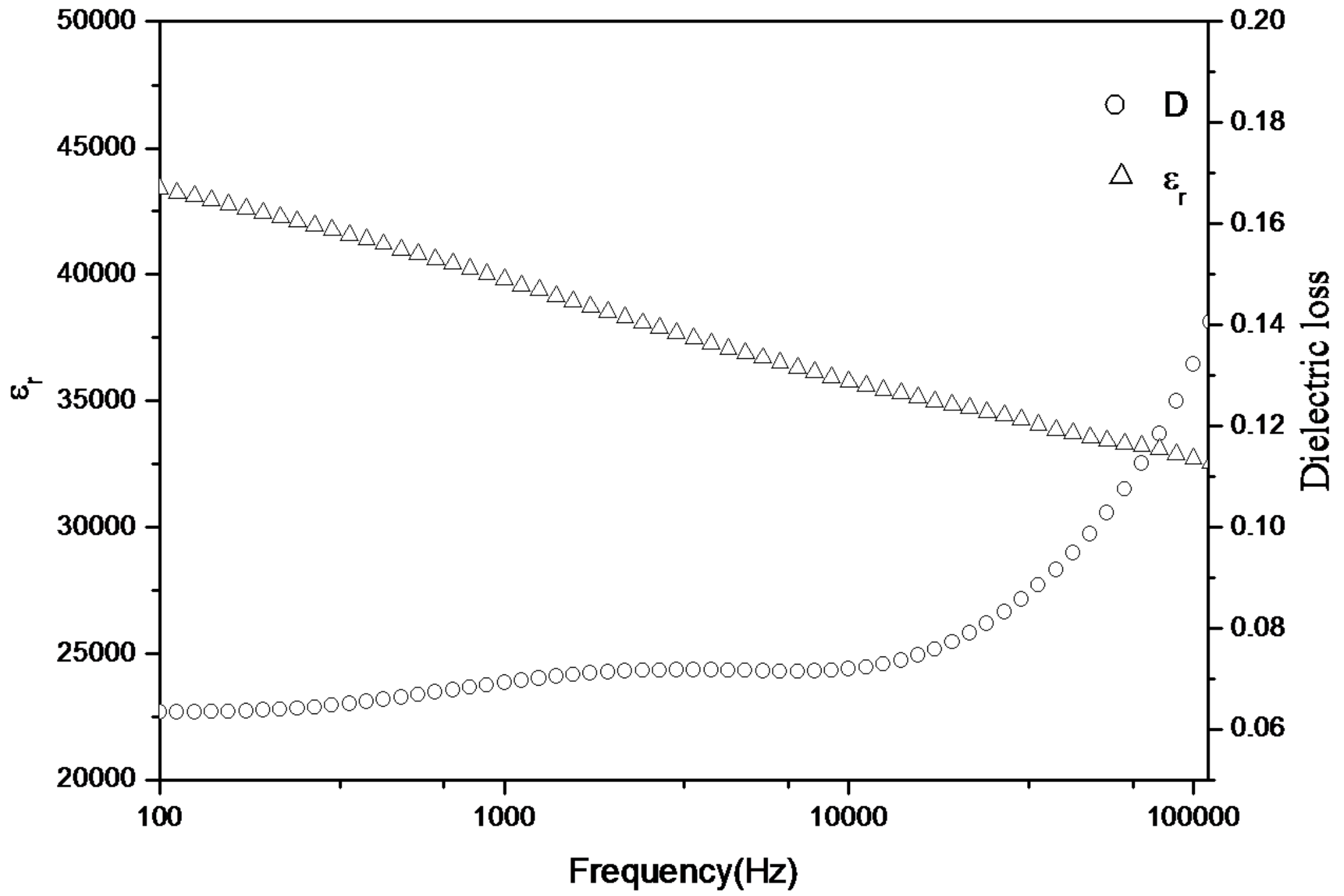

**Scheme . I.**

A clear solution (titani gel+oxalic acid) obtained on standing for several hours, according to the following reactions.

$TiO_2 .xH_2O + H_2C_2O_4 = TiO(C_2O_4)(aq)+(x+1)\ H_2O \quad (92<x<118)$ → 1

$TiO_2 \bullet xH_2O + 2H_2C_2O_4 = H_2TiO(C_2O_4)_2(aq)+(x+1)\ H_2O$ → 2

Following reaction schemes have been attempted using the clear solution.

Reaction Scheme: A

$4TiO(C_2O_4)(aq)+CaCl_2(aq)+3CuCl_2(aq)+ 4H_2C_2O_4(aq)$

↓ pH>7.0,NaOH/$NH_4OH$(aq)

$CaCu_3(TiO)_4(C_2O_4)_8$ → 3

Reaction Scheme: B

$4H_2TiO(C_2O_4)_2(aq)+CaCl_2(aq)+3CuCl_2$

↓ Solvothermal ethanol

$CaCu_3(TiO)_4(C_2O_4)_8\ nH_2O\downarrow +8HCl$ → 4

Reaction Scheme: C

$H_2TiO(C_2O_4)_2(aq)+CaCO_3 \rightarrow Ca(TiO)(C_2O_4)_2(aq)+H_2O+CO_2\uparrow$ → 5

$3H_2TiO(C_2O_4)_2(aq) + Ca(TiO)(C_2O_4)_2(aq) +3CuCO_3 \bullet Cu(OH)_2$

↓

$CaCu_3(TiO)_4(C_2O_4)_8\ nH_2O\downarrow$ → 6

Reaction Scheme: D

$3H_2TiO(C_2O_4)_2(aq)+ Ca(TiO)(C_2O_4)_2(aq) + 3[CuCl_2\ /Cu(NO_3)_2\ ] \bullet 2H_2O$

↓ Acetone+water(80:20)

$CaCu_3(TiO)_4(C_2O_4)_8\ 9H_2O\downarrow+6HCl(aq)$ → 7

**Scheme. II**

**Decomposition Scheme A (major process ) :**

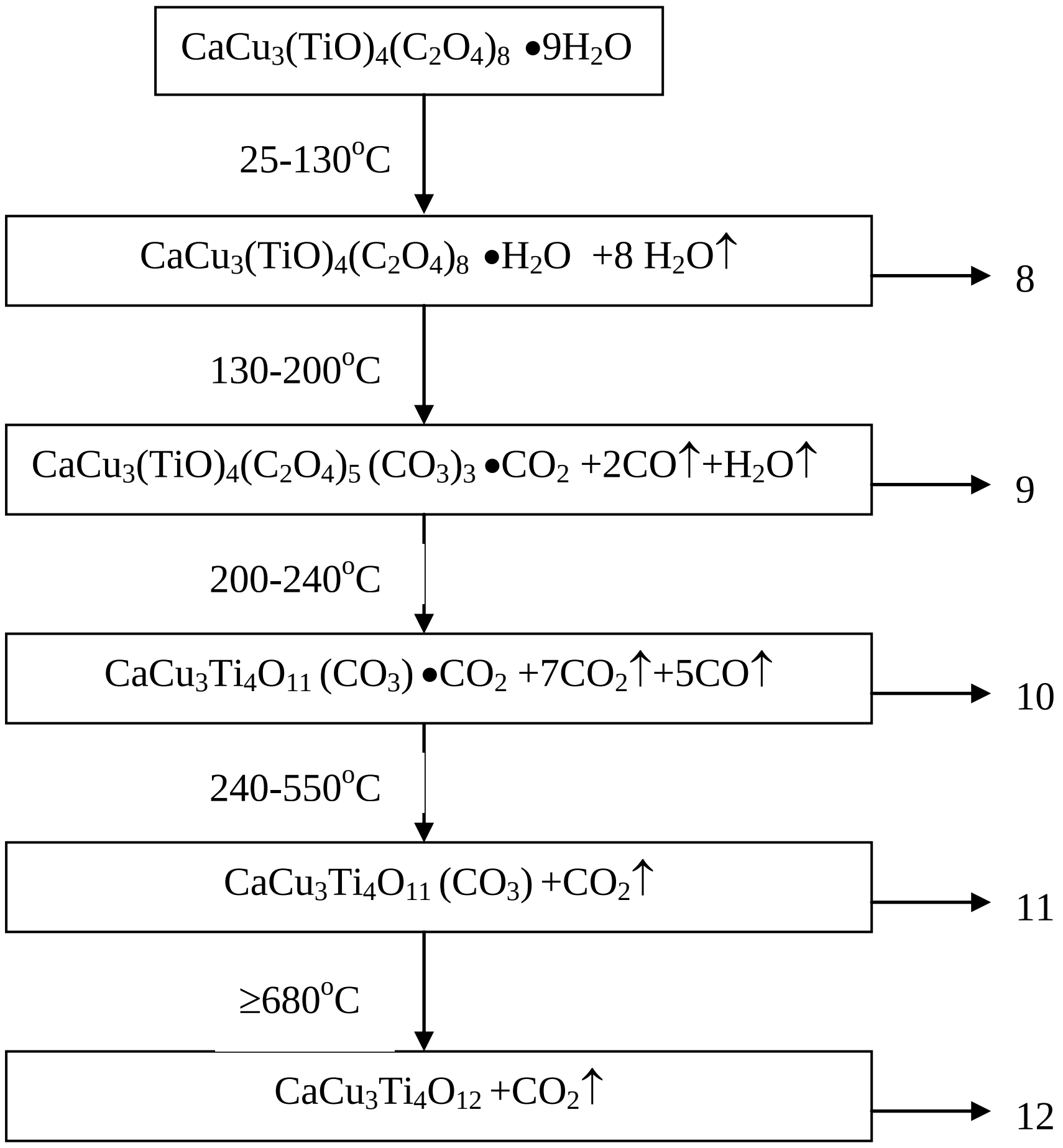

**Decomposition Scheme B ( minor process )**

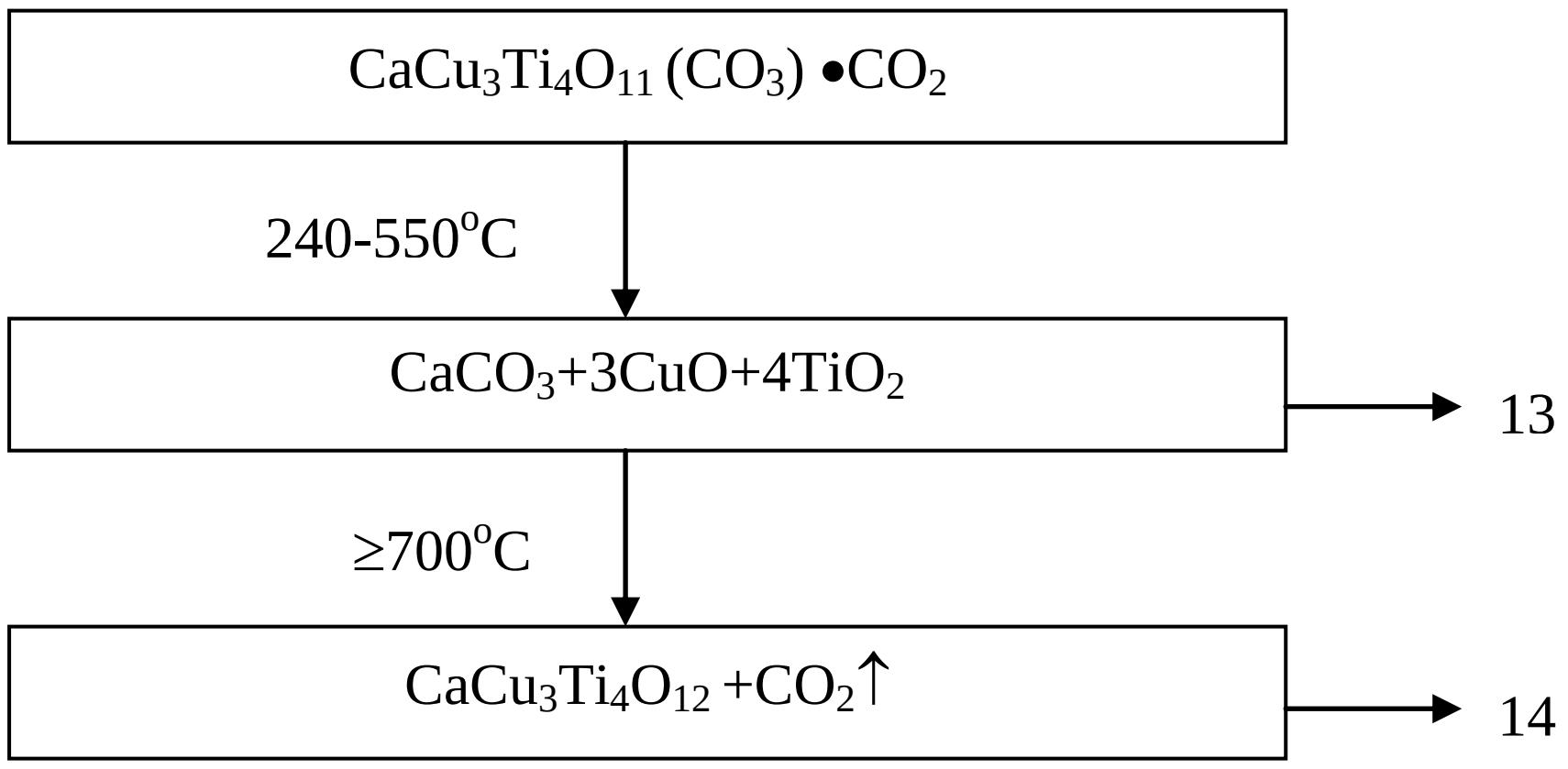